\documentclass[letterpaper,twocolumn,prl,aps,superscriptaddress,amsmath,amssymb,floatfix]{revtex4-1}
\usepackage{mathptmx}
\usepackage[latin9]{inputenc}
\usepackage{amsmath}
\usepackage{amssymb}
\usepackage{graphicx}
\usepackage{esint}

\makeatletter

\usepackage{textcomp}
\usepackage{epstopdf}

\usepackage{verbatim}

\pdfpageheight\paperheight
\pdfpagewidth\paperwidth

\@ifundefined{textcolor}{}{%
 \definecolor{BLACK}{gray}{0}
 \definecolor{WHITE}{gray}{1}
 \definecolor{RED}{rgb}{1,0,0}
 \definecolor{GREEN}{rgb}{0,1,0}
 \definecolor{BLUE}{rgb}{0,0,1}
 \definecolor{CYAN}{cmyk}{1,0,0,0}
 \definecolor{MAGENTA}{cmyk}{0,1,0,0}
 \definecolor{YELLOW}{cmyk}{0,0,1,0}
}

\usepackage{xcolor}\usepackage{soul}
\newcommand{\ket}[1]{\ensuremath{\left|#1\right\rangle}}

\definecolor{blue}{rgb}{0,0,1}
\definecolor{red}{rgb}{1,0,0}
\definecolor{green}{rgb}{0,1,0}

\makeatother

\begin{document}

\title{Experimental repetitive quantum channel simulation}

\author{L.~Hu}

\thanks{These two authors contributed equally to this work.}

\affiliation{Center for Quantum Information, Institute for Interdisciplinary Information
Sciences, Tsinghua University, Beijing 100084, China}

\author{X.~Mu}

\thanks{These two authors contributed equally to this work.}

\affiliation{Center for Quantum Information, Institute for Interdisciplinary Information
Sciences, Tsinghua University, Beijing 100084, China}

\author{W.~Cai}

\affiliation{Center for Quantum Information, Institute for Interdisciplinary Information
Sciences, Tsinghua University, Beijing 100084, China}

\author{Y.~Ma}

\affiliation{Center for Quantum Information, Institute for Interdisciplinary Information
Sciences, Tsinghua University, Beijing 100084, China}

\author{Y.~Xu}

\affiliation{Center for Quantum Information, Institute for Interdisciplinary Information
Sciences, Tsinghua University, Beijing 100084, China}

\author{H.~Wang}

\affiliation{Center for Quantum Information, Institute for Interdisciplinary Information
Sciences, Tsinghua University, Beijing 100084, China}

\author{Y.~P.~Song}

\affiliation{Center for Quantum Information, Institute for Interdisciplinary Information
Sciences, Tsinghua University, Beijing 100084, China}

\author{C.-L.~Zou}
\email{clzou321@ustc.edu.cn}
\affiliation{Key Laboratory of Quantum Information, CAS, University of Science
and Technology of China, Hefei, Anhui 230026, P. R. China}

\author{L.~Sun}
\email{luyansun@tsinghua.edu.cn}
\affiliation{Center for Quantum Information, Institute for Interdisciplinary Information
Sciences, Tsinghua University, Beijing 100084, China}

\begin{abstract}
Universal control of quantum systems is a major goal to be achieved for quantum information processing, which demands thorough understanding of fundamental quantum mechanics and promises applications of quantum technologies. So far, most studies concentrate on ideally isolated quantum systems governed by unitary evolutions, while practical quantum systems are open and described by quantum channels due to their inevitable coupling to environment. Here, we experimentally simulate arbitrary quantum channels for an open quantum system, i.e. a single photonic qubit in a superconducting quantum circuit. The arbitrary channel simulation is achieved with minimum resource of only one ancilla qubit and measurement-based adaptive control. By repetitively implementing the quantum channel simulation, we realize an arbitrary Liouvillian for a continuous evolution of an open quantum system for the first time. Our experiment provides not only a testbed for understanding quantum noise and decoherence, but also a powerful tool for full control of practical open quantum systems.
\end{abstract}
\maketitle

\section{Introduction}
Experimental quantum systems are open in reality, since they are inevitably coupled to environment. Therefore, the real physical effect on a quantum state should be a completely positive and trace-preserving mapping, also called
quantum channel, instead of a unitary evolution for an ideally isolated
quantum system~\cite{Nielsen,Petruccione2002,Wilde2013}. Therefore,
the simulation of an arbitrary quantum channel not only is of fundamental
importance for the understanding of quantum noise and decoherence,
but also allows for applications which rely on universal control of open quantum systems.
For example, quantum channel could be used for the preparation
of arbitrarily mixed quantum state, which would be a resource for fundamental
quantum mechanics and for deterministic quantum computation with one
qubit~\cite{Knill1998,Wang2018}. A continuous quantum channel can be utilized for realizing quantum state stabilization~\cite{Verstraete2009}
and autonomous quantum error correction via the quantum Zeno effect~\cite{Shankar2013,Reiter2017,Touzard2018}.
There are also appealing proposals using quantum channels for studying non-equilibrium quantum systems~\cite{Eisert2015,Neill2016}
and preparing topological quantum many-body states~\cite{Diehl2011}.

For an open quantum system constantly coupled to its environment,
the quantum state evolution follows a Liouvillian of master equation
\cite{Petruccione2002,gardiner2004quantum}. A natural choice to simulate
such a quantum channel is to construct an artificial environment and
engineer proper system-environment interaction~\cite{Albert2016,Zanardi2016}.
For example, the damping channel of a qubit can be realized by coupling
it to a resonator with low quality factor~\cite{Bienfait2016}. Rather
than such an analog approach, it is more attractive to realize a
universal digital quantum channel simulation (QCS) through a standard quantum
circuit with continuous Liouvillians which can be realized piecewisely
by repeating a quantum channel. The digital QCS
can be straightforwardly implemented based on Stinespring's dilation
with a unitary operation on the expanded Hilbert space including both the target
quantum system and the environment followed by discarding the environment
in the end. Although proof-of-principle experiments have been reported
in various quantum computing platforms~\cite{Barreiro2011,Schindler2013},
such an approach is not scalable since the required dimension of the
ancilla and number of multi-qubit gate operations scale polynomially
with the target system dimension~\cite{Terhal1999}. Recently, a convenient
approach to realize arbitrary qubit channel with minimum ancillary resource
is proposed in Ref.~\onlinecite{Wang2013}, which relies on decomposing
the channel into convex combination of quasiextreme channels. Such an approach has been demonstrated experimentally~\cite{Lu2017,McCutcheon2018}.
However, their simulation is probabilistic, and repetitive implementation
is very challenging.

\begin{figure*}
\includegraphics{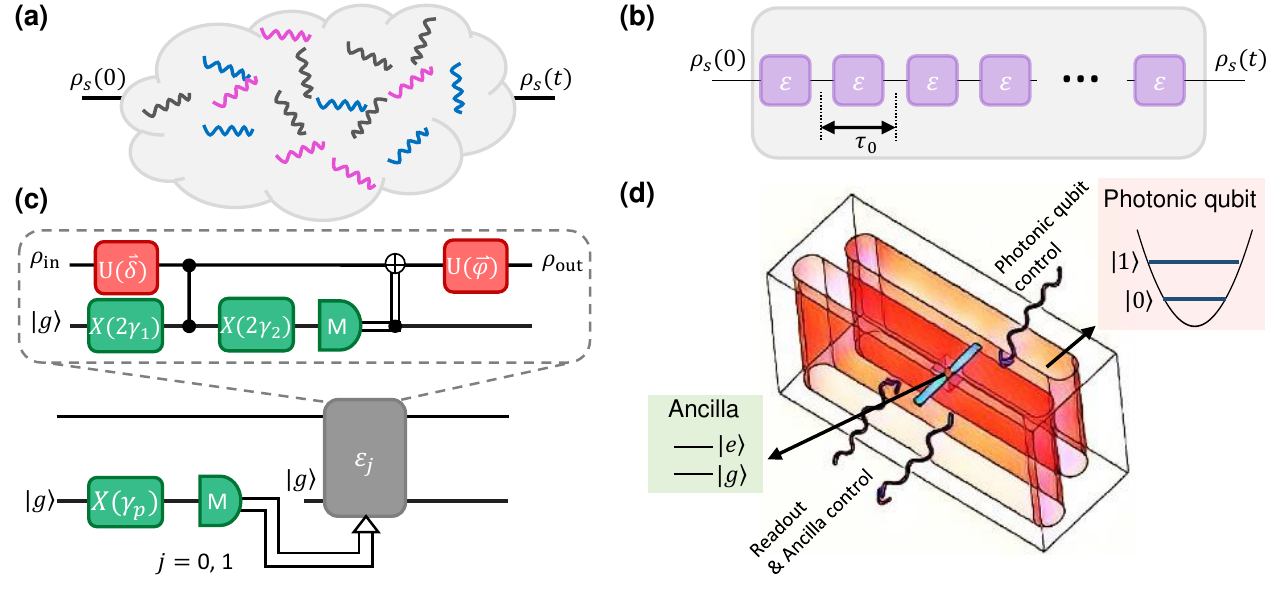} \caption{\textbf{The schematic of quantum channel simulation.} (a) Evolution
of an open quantum system coupled to a bath of harmonic oscillators.
(b) Evolution of a quantum system under repetitive quantum channel
operation $\epsilon$ with a time interval $\tau_{0}$. (c) Quantum
circuit for realizing arbitrary single-qubit quantum channels with
the assistance of one ancilla qubit. $U$ is a unitary operation on the system qubit. $X(\gamma)$ represents a rotation of the ancilla qubit along $X$-axis with an angle $\gamma$. $M$ denotes a measurement on the ancilla qubit. (d) Experimental system of a three-dimensional cQED architecture for
repetitive QCS. Fock states \{$\ket{0}$, $\ket{1}$\}
in a microwave cavity constitute the photonic qubit on which channel
simulations are performed, while a transmon qubit with states $\left\{ \left|g\right\rangle ,\,\left|e\right\rangle \right\} $
serves as an ancilla.}
\label{fig:Fig1} \vspace{-6pt}
\end{figure*}

Here we deterministically implement repetitive single-qubit digital QCS within a superconducting system with a circuit quantum electrodynamics (cQED) architecture~\cite{Wallraff,Paik2011,Devoret2013}. The digital QCS benefits from the high-fidelity quantum non-demolition (QND) measurement of superconducting qubits and fast real-time adaptive control based on field programmable gate arrays (FPGA). We first realize two typical channels of dephasing and amplitude damping for a system qubit, and demonstrate the controlling of external dephasing and damping rates in a large range (2-3 orders of magnitude) respectively. Furthermore, we realize arbitrary QCS based on the proposal in Ref.~\onlinecite{Wang2013}, which yields an average state generation fidelity of 97\% and is mainly limited by the decoherence of the ancilla qubit. The repetitive QCS with controllable parameters provides a testbed for studying reservoir engineering~\cite{Verstraete2009}, quantum Zeno effect~\cite{Shankar2013,Reiter2017,Touzard2018,Ficheux2018}, quantum thermodynamics~\cite{Vinjanampathy2016,Alicki2018}, and quantum metrology~\cite{Laurenza2018,Yuan2017}. Together with the recently demonstrated arbitrary unitary control and quantum error correction~\cite{Ofek2016,Hu2018} in cQED architecture, our demonstrated QCS could realize reliable universal control of open quantum system, which is significant for quantum computation~\cite{Nielsen,Devoret2013} and simulation~\cite{Houck2012,Salathe2015}.

\vspace{-10pt}
\section{Results}

\subsection{Principle and system}

Figure$\,$\ref{fig:Fig1}(a) is a schematic of the time evolution
of an open quantum system coupled to a reservoir of harmonic oscillators.
Due to the inaccessibility of the environment, the evolution of the
system is described by a quantum channel $\mathcal{E}_{t}:\,\rho_{\mathrm{s}}\left(0\right)\mapsto\rho_{\mathrm{s}}\left(t\right)$,
instead of unitary operations for a closed quantum system. For
a typical open quantum system with time-independent system-environment
interaction, the Hamiltonian under the rotating-wave approximation
reads $H_{\mathrm{se}}=\int d\omega\left(g_{\omega}o^{\dagger}b_{\omega}+h.c.\right)$,
where $b_{\omega}$ is the bosonic operator associated with the environment
degrees of freedom, $o$ is the jump operator of the system, and $g_{\omega}$
is the interaction strength. Applying the Born-Markov approximation,
the evolution of the system or the quantum channel follows the master equation, and the dynamics
of quantum states satisfy {[}Supplementary Information{]}
\begin{equation}
\rho_{\mathrm{s}}\left(t\right)=e^{t\mathcal{L}}\rho_{\mathrm{s}}\left(0\right),
\end{equation}
with the Lindblad form of the Liouvillian as \cite{gardiner2004quantum}
\begin{equation}
\mathcal{L}\rho=\sum\kappa\left(2o\rho o^{\dagger}-o^{\dagger}o\rho-\rho o^{\dagger}o\right).\label{eq:Liouvillian}
\end{equation}
Here, $\kappa$ is the decoherence rate determined by $g_{\omega}$
and the density of states of the environment.

Alternatively, the quantum channel can be represented in the Kraus
representation as \cite{Nielsen}
\begin{equation}
\rho_{\mathrm{s}}\left(t\right)=\mathcal{E}_{t}\left[\rho_{\mathrm{s}}\left(0\right)\right]=\sum_{k}E_{k}^{\left(t\right)}\rho_{\mathrm{s}}\left(0\right)E_{k}^{\left(t\right)\dagger},
\end{equation}
with the Kraus operators satisfying $\sum_{k}E_{k}^{\left(t\right)\dagger}E_{k}^{\left(t\right)}=\mathbb{I}$.
This formula describes a discrete mapping between quantum states at
different times.

\begin{figure*}
\includegraphics[width=0.75\textwidth]{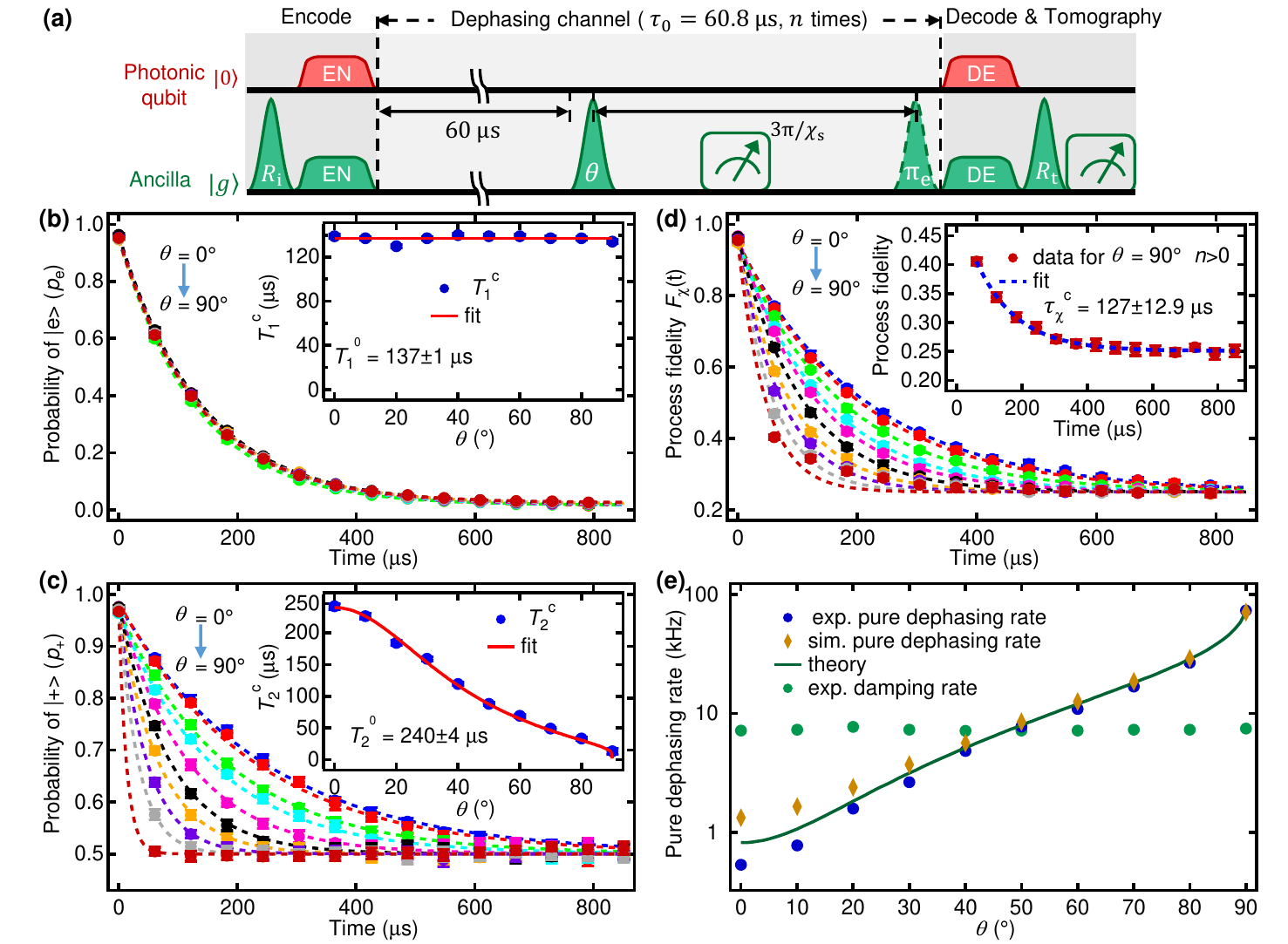}\caption{\textbf{Repetitive single-qubit dephasing channel simulation.} \textbf{(a)}
Experimental sequence. In each experiment, the initialization and
encoding processes are performed at the very beginning, and the quantum
channel (consisting of two operations on the ancilla qubit and a controlled-phase
gate between the ancilla and the photonic qubit over a time interval
of $3\pi/\chi_{{\rm {s}}}$) is repeated $n$ times. Decoding and
tomography are performed in the end to characterize the channel performance.
\textbf{(b)} and \textbf{(c) }The longitudinal and transverse relaxation
of the photonic qubit with repetitive QCS.
Insets: the extracted $T_{1}^{{\rm {c}}}$ and $T_{2}^{{\rm {c}}}$
against the control parameter $\theta$ of the channel. $T_{1}^{{\rm {c}}}$
curves are fitted with $1/{T_{1}^{{\rm {c}}}}=1/{{T_{1}}^{0}}$ while
$T_{2}^{{\rm {c}}}$ curves are fitted with $1/{T_{2}^{{\rm {c}}}}=-{\rm ln}([2{\rm cos}^{2}(\theta/2)-1])/\tau_{0}+1/{{T_{2}}^{0}}$,
where $T_{1}^{0}$ and $T_{2}^{0}$ are free parameters and approach the intrinsic decoherence times $T_1^{\mathrm{s}}$ and $T_2^{\mathrm{s}}$ of the photonic qubit. \textbf{ (d)}
The decay of channel process fidelity $F_{\chi}$ with different $\theta$.
\textbf{(e)} The pure dephasing rate is controlled over a range of
two orders of magnitude by varying $\theta$, in good agreement with
theoretical expectation and simulation.}

\label{fig:dephasing} \vspace{-6pt}
\end{figure*}

According to the two descriptions of quantum channels, there are different
approaches to simulate a quantum channel. It is straightforward to
simulate the continuous dynamics of an open quantum system through an analog approach, in which a given Liouvillian can be realized by directly constructing an environment with certain density of states and engineering the proper
system-environment interaction. Another approach is piecewise implementation of a channel $\mathcal{E}_{\tau_{0}}$ by a standard quantum circuit
with an interval $\tau_{0}$, as shown in Fig.$\,$\ref{fig:Fig1}(b).
The dynamics of the open quantum system can be digitally simulated
as $\mathcal{E}_{t}=\left[\mathcal{E}_{\tau_{0}}\right]^{t/\tau_{0}}:\,\rho_{\mathrm{s}}\left(0\right)\mapsto\rho_{\mathrm{s}}\left(t\right)$.
This digital approach is universal because arbitrary continuous evolution of a quantum
system in time can be realized by repetitively implementing QCS. For instance, the Liouvillian of Eq.$\,$(\ref{eq:Liouvillian})
can be simulated digitally in the limit of $\kappa\tau_{0}\ll1$ {(}Supplementary Information, also see~ \cite{Andersson2007}{)}, with
\begin{eqnarray}
E_{0}^{\left(\tau_{0}\right)} & = & \sqrt{2\kappa\tau_{0}}o,\\
E_{1}^{\left(\tau_{0}\right)} & = & \mathbb{I}-\kappa\tau_{0}o^{\dagger}o.
\end{eqnarray}

Although the Markovian environment contains huge degrees of freedom, as proved theoretically~\cite{Wang2013,Shen2017} an arbitrary quantum channel for a system qubit can be
efficiently and deterministically simulated with minimum resource of one single ancilla qubit and measurement-based adaptive control. The quantum circuit
for an arbitrary qubit QCS is shown in Fig.$\,$\ref{fig:Fig1}(c).
As a ``quantum dice'', when the ancilla qubit (initially in the
ground state $\ket{g}$) is tossed by rotating it along $X$-axis
with an angle $\gamma_{p}$ and then measuring it in the computational
basis $\left\{ \left|g\right\rangle ,\,\left|e\right\rangle \right\} $,
a random bit $j\in\left\{ 0,\,1\right\} $ is generated with the probabilities
$\left\{ \cos^{2}\frac{\gamma_{p}}{2},\,\sin^{2}\frac{\gamma_{p}}{2}\right\} $.
By randomly choosing the quantum circuit $\mathcal{E}_{j}$ (a quasiextreme
channel) according to the result of the toss, an arbitrary qubit channel
can be obtained deterministically as a convex combination of these
two quasiextreme channels~\cite{Wang2013}. After the toss, the ancilla qubit is also
utilized in the quasiextreme channel simulations {[}Fig.$\,$\ref{fig:Fig1}(c){]}
to assist the unitary operations on the system qubit, e.g.,
implementing a controlled-phase (CZ) gate and measurement-based adaptive
control as in our experiments.

As schematically shown in Fig.$\,$\ref{fig:Fig1}(d), our experimental
device consists of a superconducting transmon qubit dispersively coupled
to two waveguide cavity resonators~\cite{Paik2011,Kirchmair,Vlastakis2013,SunNature,Liu2017,Hu2018}.
One of the cavities (storage cavity) has long photon coherence times
$T_{1}^{\mathrm{s}}=143~{\mathrm{\mu}}$s and $T_{2}^{\mathrm{s}}=250~{\mathrm{\mu}}$s,
and its $\ket{0}$ and $\ket{1}$ Fock states constitute the two bases
of a photonic qubit (the system qubit) on which the QCS are performed. The transmon qubit as an ancilla with an
energy relaxation time $T_{1}=30~{\mathrm{\mu}}$s and a pure dephasing time
$T_{\varphi}=120~{\mathrm{\mu}}$s is used to realize the QCS.
The other short-lived cavity with a photon lifetime $\tau_{{\rm {r}}}=44$~ns
is to readout the ancilla qubit with the help of a phase-sensitive
Josephson bifurcation amplifier~\cite{Hatridge,Roy2015,Kamal,Murch}
for a high fidelity single-shot QND measurement. Each readout
measurement throughout our experiment returns a digitized value of
the qubit state (0 and 1 correspond to the ground state $\ket{g}$ and excited state $\ket{e}$, respectively). Fast real-time adaptive control is also vital for the demonstrated QCS. This is achieved through
three FPGAs with home-made logics, which allow us to individually control the ancilla qubit, the photonic
qubit, and the readout cavity, and also integrate readout signal sampling, ancilla state estimation,
and manipulation signal generation together. The experimental apparatus
and readout properties are similar to the earlier report in Ref.~\onlinecite{Hu2018}.

The ancilla qubit and storage cavity are well described by the dispersive
Hamiltonian (The readout cavity has been neglected since it remains
in vacuum unless a measurement is performed)
\begin{equation}
H/\hbar=\omega_{s}a^{\dagger}a+\omega_{a}|e\rangle\langle e|-\chi_{{\rm {s}}}a^{\dagger}a|e\rangle\langle e|\label{eq:Hamiltonian}
\end{equation}
where $a^{\dagger}(a)$ is the creation (annihilation) operator of
the storage cavity, $|e\rangle$ is the excited state of the ancilla
qubit, and $\chi_{{\rm {s}}}/2\pi=1.90$~MHz is the dispersive interaction
strength between the qubit and the storage cavity. This strong dispersive
coupling gives rise to the necessary operations for the simulation,
including unitary operations on the photonic qubit and the two-qubit
gates. The control pulses are numerically optimized with the gradient
ascent pulse engineering (GRAPE) method~\cite{Khaneja2005,DeFouquieres2011}
based on carefully calibrated experimental parameters.

\begin{figure*}
\includegraphics[width=0.75\textwidth]{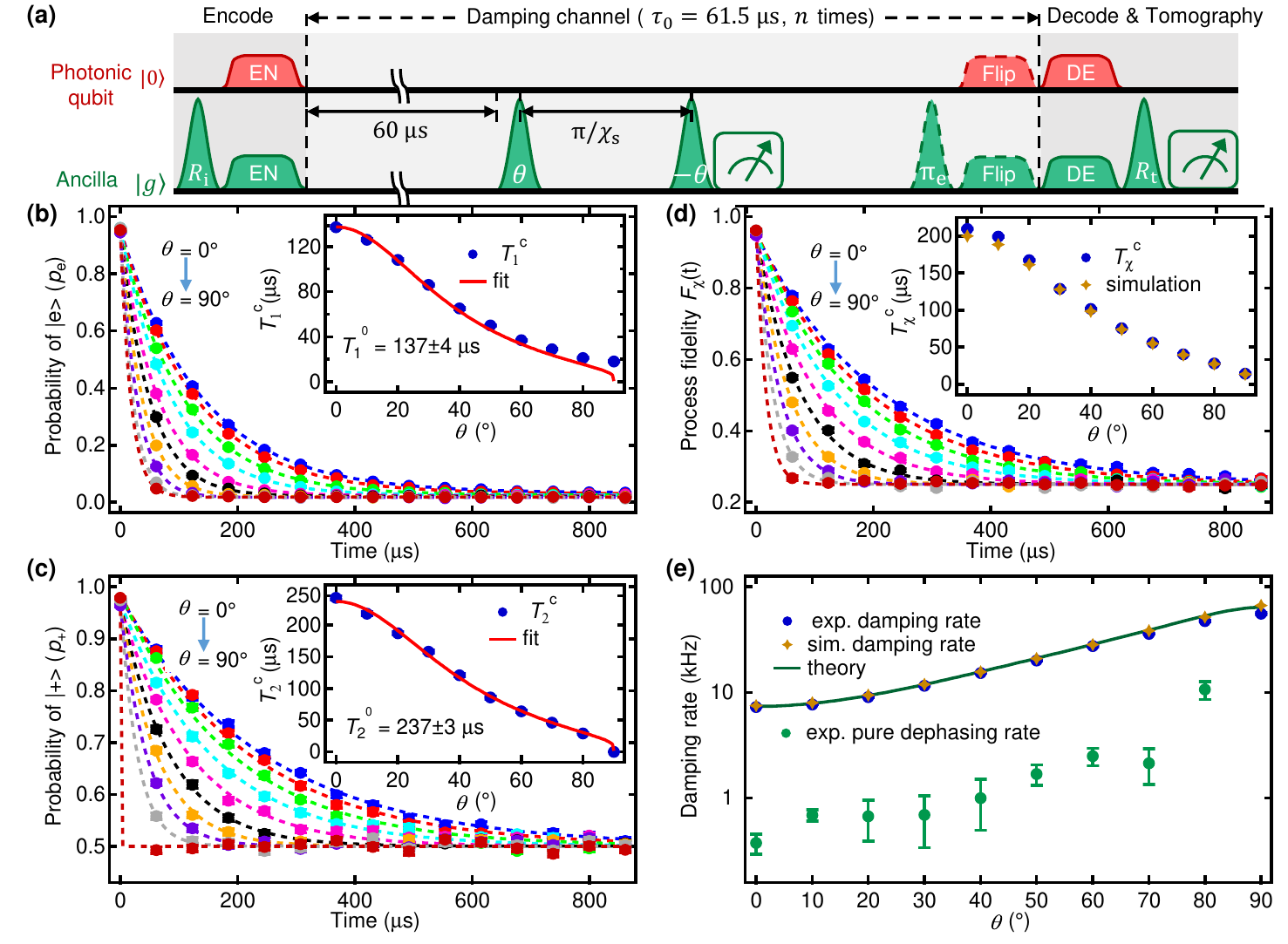}\caption{\textbf{Repetitive single-qubit damping channel simulation.} \textbf{(a)}
Experimental sequence. \textbf{(b) }and\textbf{ (c) }The longitudinal
and transverse relaxation of the photonic qubit with repetitive QCS. Insets: the extracted $T_{1}^{{\rm {c}}}$ and
$T_{2}^{{\rm {c}}}$ against the control parameter $\theta$ of the
channel. $T_{1}^{{\rm {c}}}$ curves are fitted with $1/{T_{1}^{{\rm {c}}}}=-{\rm ln}({\rm cos}^{2}\theta)/\tau_{0}+1/{{T_{1}}^{0}}$,
resulting in ${{T_{1}}^{0}}=137~{\mathrm{\mu}}$s. $T_{2}^{{\rm {c}}}$ curves
are fitted with $1/{T_{2}^{{\rm {c}}}}=-{\rm ln}({\rm cos}^{2}\theta)/2\tau_{0}+1/{{T_{2}}^{0}}$,
resulting in ${{T_{2}}^{0}}=237~{\mathrm{\mu}}$s. \textbf{(d)} The decay
of channel process fidelity $F_{\chi}$ with different $\theta$. \textbf{(e)}
The damping and pure dephasing rates versus $\theta$, obtained from
experimental results, simulation, and theoretical prediction, respectively.}

\label{fig:damping} \vspace{-6pt}

\end{figure*}

\begin{figure*}
\includegraphics[width=0.8\textwidth]{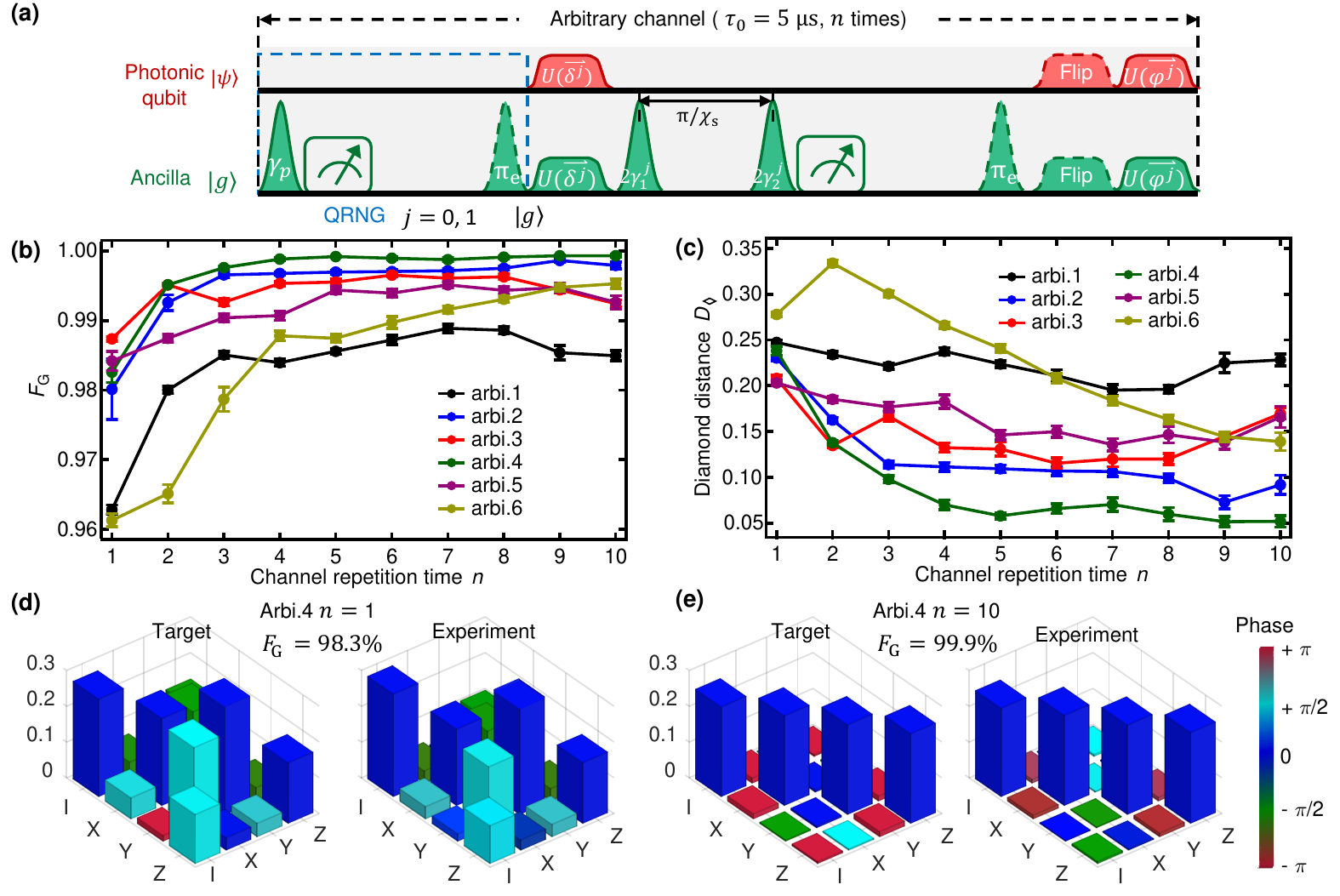} \caption{\textbf{Arbitrary single-qubit quantum channel simulation.} \textbf{(a)} Pulse sequence for the arbitrary QCS protocol shown in Fig.$\,$\ref{fig:Fig1}(c). The QCS starts with a quantum random number generation (QRNG), realized by measuring the ancilla in a superposition state. The channel is repeated $n=1\sim10$
times without any time delay. \textbf{(b)} State generation fidelity $F_{\mathrm{G}}$
between the experimental and theoretical $\chi$ matrices as a function
of $n$ for six different arbitrary channels. The simulated arbitrary
channels at $n=1$ yield an average fidelity of $97\%$. The error
bars are determined by bootstrapping on the measured $\chi$ matrices.
\textbf{(c)} Calculated diamond distance $D_{\diamond}$ between the experimental
$\chi_{{\rm {E}}}$ and the target $\chi_{{\rm {T}}}$ as a function
$n$ for the six arbitrary channels. \textbf{(d)} and \textbf{(e)}
Typical $\chi_{\mathrm{E}}$ and $\chi_{\mathrm{T}}$ of
the 4th channel at $n=1$ and $n=10$, respectively.
The height and color represent the amplitude and phase, respectively.
The experiment results are in excellent agreement with theoretical
expectations.}
\label{fig:arbitrary} \vspace{-6pt}

\end{figure*}

\subsection{Qubit dephasing channel}

To illustrate the concept of ancilla-assisted qubit QCS,
we start with the simple dephasing channel simulation for the photonic
qubit. As shown by the experimental pulse sequence in Fig.~\ref{fig:dephasing}(a),
the system is initialized to $\left|0\right\rangle \otimes\left|g\right\rangle $,
and the photonic qubit is prepared to a pure state with the assistance
of the ancilla, while the ancilla goes back to $\ket{g}$ after the
encoding process. Then, the dephasing channel is implemented through
three steps: (i) state preparation of the ancilla qubit by $X_{\theta}$,
i.e., a rotation of the ancilla with an angle $\theta$ along the
$X$-axis. (ii) CZ gate induced by the dispersive interaction between
the ancilla and the photonic qubit over a time interval of $3\pi/\chi_{\mathrm{s}}$.
(iii) a projective measurement followed by a conditional gate on the
ancilla qubit to reset it to $\ket{g}$, thus erasing the quantum
information on the ancilla. During the channel simulation, there is
no external operation on the photonic qubit. Therefore, the population
of the photonic qubit is not changed by the channel and only the
phase of the photonic qubit is entangled with the ancilla that induces
the dephasing process. Thus, the quantum channel $\mathcal{E}_{\theta}^{\mathrm{dph}}:\,\rho_{\mathrm{s}}\mapsto p_{\theta}\rho_{\mathrm{s}}+\left(1-p_{\theta}\right)\mathbf{Z}\rho_{\mathrm{s}}\mathbf{Z}$
is realized, where $p_{\theta}=\cos^{2}\frac{\theta}{2}$ and $\mathbf{Z}$
is the Pauli matrix. The dephasing channel can be implemented deterministically
without post-selection and also without destroying the photonic qubit. The channel
simulation therefore can be repeated $n$ times with a controlled
repetition interval $\tau_{0}$ and dephasing probability $p_{\theta}$
to simulate an open quantum system with arbitrary dephasing rate $\kappa=-\ln\left(2p_{\theta}-1\right)/\tau_{0}$.
Finally, the information is decoded back from the photonic qubit to
the ancilla with a state tomography for an evaluation of the channel
performance.

In the experiment presented in Fig.~\ref{fig:dephasing}, we fix the repetition interval $\tau_{0}\approx61~{\mathrm{\mu}}$s
while changing $\theta$, i.e., the dephasing rate continuously. Figures~\ref{fig:dephasing}(b)
and \ref{fig:dephasing}(c) show the measured probability $p_{e}=\left\langle e\right|\left(\mathcal{E}_{\theta}^{\mathrm{dph}}\right)^{n}\left(\rho_{\mathrm{in}}\right)\left|e\right\rangle $
and $p_{+}=\left\langle +\right|\left(\mathcal{E}_{\theta}^{\mathrm{dph}}\right)^{n}\left(\rho_{\mathrm{in}}\right)\left|+\right\rangle $
with the initial state of $\rho_{\mathrm{in}}=\left|1\right\rangle \left\langle 1\right|$
and $\left|+\right\rangle \left\langle +\right|$, respectively. Here,
$\ket{+}=\left(\ket{g}+\ket{e}\right)/\sqrt{2}$. The results correspond
to the longitudinal and transverse relaxation of the photonic qubit
induced by the channel, from which we can extract $T_{1}^{\mathrm{c}}$
and $T_{2}^{\mathrm{c}}$ of the simulated channel coherence times
for different $\theta$. As expected, $T_{1}^{\mathrm{c}}$ is almost
not affected by this dephasing channel while the effective dephasing
time $T_{2}^{\mathrm{c}}$ becomes shorter for a larger $\theta$.
When $\theta=0$ (no extra dephasing), we get ${{T_{1}^{\mathrm{c}}}(0)}=139~{\mathrm{\mu}}$s
and ${{T_{2}^{\mathrm{c}}}(0)}=242~{\mathrm{\mu}}$s, agreeing well with the
intrinsic coherence times of the photonic qubit.

To fully characterize the channel, we also measure the process $\chi$
matrix of the channel, and calculate the process fidelity $F_{\chi}(t)=\mathrm{Tr}\left(\chi_{\mathrm{dph}}\chi_{\mathrm{I}}\right)$
with $\chi_{\mathrm{dph}}$ and $\chi_{\mathrm{I}}$ being the $\chi$
matrix of the measured dephasing channel and the identity channel,
respectively. As shown in Fig.~\ref{fig:dephasing}(d), the first
point $F_{\chi}(0)=95.5\%$ indicates a high ``round trip\char`\"{}
process fidelity of the encoding and decoding processes only. Due
to the large dephasing rate when $\theta$ approaches $90^{\circ}$,
the photonic qubit is nearly completely dephased after the first round of QCS. This should result in $F_{\chi}\approx0.5$. However, due
to the intrinsic decoherence of the photonic qubit, $F_{\chi}$ is lower
than the limit of $0.5$ and eventually converges to the limit of
$0.25$ for a damping channel. To confirm this, a fit to the results
for $n\ge1$ ($t>0$) at $\theta=90^{\circ}$ is shown in Fig.~\ref{fig:dephasing}(d)
inset, which gives a decay of $127~{\mathrm{\mu}}$s agreeing well with the intrinsic
photon ${T_{1}}^{{\rm s}}$.

Figure~\ref{fig:dephasing}(e) shows
the experimental pure dephasing rate ($1/T_{2}^{{\rm {c}}}-1/2T_{1}^{{\rm {c}}}$)
as a function of $\theta$, in excellent agreement with theoretical
expectation (see Supplementary Information) and simulation based on
QuTiP in Python~\cite{Johansson2012,Johansson2013} (both take the ancilla decoherence into account). As demonstrated in Fig.~\ref{fig:dephasing}(e), the pure dephasing rate can be controlled over a range of two orders of magnitude by varying $\theta$. Since our QCS can be implemented within a duration of $\mathrm{\mu}$s, the channel repetition interval can be much shorter than $61~\mathrm{\mu}$s used here. The experimental results obtained with $\tau_0=5.8~\mathrm{\mu}$s are presented in Supplementary Information, and we can achieve a pure dephasing rate more than 1000 times faster than the intrinsic pure dephasing rate.

\subsection{Qubit amplitude damping channel}

We next implement a single-qubit amplitude damping channel, which is also a
quasiextreme channel {[}Fig.$\,$\ref{fig:Fig1}(c){]}. The pulse
sequence is shown in {[}Fig.~\ref{fig:damping}(a){]}, with the same
initialization, encoding, and decoding processes as for the dephasing
channel. Here, the kernel part of the damping channel is realized
through a Ramsey-type of measurement with an interval of $\pi/\chi_{{\rm {s}}}$
between two rotations $X_{\theta}$ and $X_{-\theta}$. In contrast
to the dephasing channel, adaptive control on the photonic qubit is
necessary. If the ancilla collapses onto $\left|e\right\rangle $
after a detection, a GRAPE pulse is used to flip the photonic qubit
($\mathbf{X}$ gate) following a $\pi$ pulse to first reset the ancilla.
The rotation angle $\theta$ determines the probability $p_{\theta}=\sin^{2}\theta$
of the photonic qubit being flipped to $\left|0\right\rangle $ by
the conditional $\mathbf{X}$ gate. Therefore, the quantum channel
acting on the photonic qubit is a damping channel $\mathcal{E}_{\theta}^{\mathrm{dmp}}:\,\rho_{\mathrm{s}}\mapsto E_{0}\rho_{\mathrm{s}}E_{0}^{\dagger},+E_{1}\rho_{\mathrm{s}}E_{1}^{\dagger}$,
with $E_{0}=\left|0\right\rangle \left\langle 0\right|+\sqrt{1-p_{\theta}}\left|1\right\rangle \left\langle 1\right|$
and $E_{1}=\sqrt{p_{\theta}}\left|0\right\rangle \left\langle 1\right|$.
By repetitively implementing the damping channel with an interval
$\tau_{0}$, an effective damping rate $\kappa=-\ln\left(1-p_{\theta}\right)/\tau_{0}$
can be realized.


To compare with the dephasing channel, we fix $\tau_{0}\approx62~{\mathrm{\mu}}$s
and vary $\theta$ to study the performance of the channel. From the
results presented in Figs.~\ref{fig:damping}(b) and \ref{fig:damping}(c), the longitudinal
and transverse relaxation times of the channel $T_{1}^{\mathrm{c}}$
and $T_{2}^{\mathrm{c}}$ are extracted and plotted in the corresponding
insets. At $\theta=0$, ${{T_{1}^{\mathrm{c}}}(0)}=137~{\mathrm{\mu}}$s and
${{T_{2}^{\mathrm{c}}}(0)}=243~{\mathrm{\mu}}$s approaching the intrinsic ones.
When increasing $\theta$, both longitudinal and transverse relaxation
times reduce, and $T_{2}^{\mathrm{c}}\approx2T_{1}^{\mathrm{c}}$
is satisfied for a damping channel with negligible intrinsic pure
dephasing process. When $\theta$ approaches $90^{\circ}$, the photonic
qubit should be completely damped to $\ket{g}$ after the first channel
simulation process. However, it is noticed that the experimental $T_{1}^{\mathrm{c}}$
saturates at certain value due to the imperfections of the ancilla
qubit. In Fig.~\ref{fig:damping}(d), the process fidelity $F_{\chi}(t)$
is also evaluated for the damping channel, which saturates to the
limit $0.25$ as expected for a damping process. Figure~\ref{fig:damping}(e)
shows the experimental damping rate as a function of $\theta$, also
in excellent agreement with theoretical expectation (see Supplementary
Information) and simulation (both take the ancilla decoherence into account). This indicates a negligible pure
dephasing effect from the simulated damping channel, also consistent
with the nearly constant pure dephasing rate extracted from experiment.


\subsection{Arbitrary quantum channel simulation}

The experimental demonstrations of the dephasing and damping channels
prove that our cQED platform and the adaptive control are reliable
for QCS. Now, we turn to the demonstration
of the most general quantum tool for arbitrary qubit QCS.
The quantum circuit in Fig.~\ref{fig:Fig1}(c) is realized with the
experimental sequence in Fig.~\ref{fig:arbitrary}(a). Here, the
encoding, decoding, and tomography procedures are the same as those
in Fig.~\ref{fig:dephasing}(a) and Fig.~\ref{fig:damping}(a),
and are omitted for simplicity. At the beginning of the QCS,
the ancilla qubit is prepared in a quantum superposition state of
$\left|g\right\rangle $ and $\left|e\right\rangle $, and a projective
measurement on it generates a quantum random number $j\in\left\{ 0,1\right\} $.
According to $j$, an adaptive operation on the ancilla initializes
the ancilla back to $\left|g\right\rangle $, and then the $j$th quasiextreme
channel is implemented. The pulse sequences for the two quasiextreme
channels are exactly the same, except that the 8 parameters $\left\{ \overrightarrow{\delta},\,\gamma_{1},\,\gamma_{2},\overrightarrow{\varphi}\right\} $
are different.

To simulate the arbitrary quantum channel, we first numerically generate
a random quantum channel $\mathcal{E}_{\mathrm{T}}$ with the target
process matrix $\chi_{\mathrm{T}}$. Using the difference between
$\chi_{\mathrm{T}}$ and the expected process matrix by the quantum
circuit in Fig.~\ref{fig:Fig1}(c) as the figure of merit, the experimental parameters
for the quasiextreme channels and the quantum random number are optimized
to minimize the difference. We then experimentally measure the process
matrix $\chi_{\mathrm{E}}$ using the optimized parameters
by repeating the experimental channel $n=1\sim10$ times without any
time delay ($\tau_0=5~{\mathrm{\mu}}$s).

To characterize the performance of the arbitrary QCS, we use the fidelity of state generation $F_{\mathrm{G}}=\inf_{\rho}\sqrt{\sqrt{\mathcal{E}_{\mathrm{T}}\left(\rho\right)}\left[\mathcal{E}_{\mathrm{E}}\left(\rho\right)\right]\sqrt{\mathcal{E}_{\mathrm{T}}\left(\rho\right)}}$,
which corresponds to the worst fidelity of the quantum state generated
by the experimental channel $\mathcal{E}_{\mathrm{E}}$ (based
on $\chi_{{\rm {E}}}$) when compared with the one by the target
channel $\mathcal{E}_{\mathrm{T}}$.
Figure~\ref{fig:arbitrary}(b) shows $F_{\mathrm{G}}$ as a function
of the channel repeated times $n$ for six different arbitrary channels.
For $n=1$, the evaluated average fidelity for the six arbitrary channels
is $97\%$. Figures~\ref{fig:arbitrary}(d) and \ref{fig:arbitrary}(e) show the typical
$\chi_{\mathrm{E}}$ and $\chi_{\mathrm{T}}$ of the 4th channel at
$n=1$ and $n=10$, respectively, demonstrating
excellent agreement between experiment and theory. As an alternative
measure, the diamond distance $D_{\diamond}$~\cite{Wilde2013} is
also applied to evaluate the performance of our experimental QCS, as depicted in Fig.$\,$\ref{fig:arbitrary}(c). The average
$D_{\diamond}$ is about $0.25$ at $n=1$. It is interesting to note
that the repetitive QCS with increased $n$ shows an
increment of fidelity or a reduction of distance. An intuitive explanation
for this behavior is that the repetitive arbitrary channel converges
to a depolarization channel when both the target and experimental
channels are with imperfections, and the difference between $\chi_{\mathrm{E}}$
and $\chi_{\mathrm{T}}$ reduces with increased $n$.



\section{Discussion}

The demonstrated repetitive QCS is important
for understanding fundamental decoherence processes in an open quantum
system. It also represents a significant step towards full control
of an open quantum system, which promises to manipulate and stabilize quantum states, as well as realize general positive-operator valued measure. Inspired by Lloyd and Viola~\cite{Lloyd2001}, it was
proved that QCS of arbitrary dimension
can be efficiently realized with a single ancilla qubit and adaptive
control~\cite{Shen2017}. Therefore, the demonstrated QCS scheme can also be generalized to a photonic qudit with
$d$-levels~\cite{Shen2017}, using exactly the same experimental setup and requiring only a single ancilla qubit and about $2\log_{2}d$ steps of adaptive control. Our work thus paves the way to arbitrary QCS of a bosonic oscillator with high-dimensional Hilbert space.

It is worth noting that our ultrafast adaptive control allows for the implementation of channels within a duration of $\mathrm{\mu s}$ and thus provides a platform for simulating continuous dynamics of
quantum systems in an artificial environment. It was
proposed in Refs.~\cite{Kliesch2011,Sweke2014} that an arbitrary
quantum channel can also be simulated with a given error bound by
the Suzuki-Lie-Trotter decomposition, similar
to the arbitrary Hamiltonian construction~\cite{Lloyd1996}. Therefore, instead of a convex combination of quasiextreme channels as demonstrated, our experimental architecture with ultrafast adaptive control is also suitable for implementing an arbitrary Liouvillian
$\mathcal{L}=\sum_{j}\mathcal{L}_{j}$ by alternatively
and piecewisely simulating its components $\mathcal{L}_{j}$ {[}Supplementary Information{]}. In addition, although our experimental results show
high fidelity of state generation by the simulated quantum channel,
further improvement calls for incorporating quantum error correction~\cite{Ofek2016,Hu2018} into this scheme.

\bibliographystyle{Zou}
%

\vspace{0.2in}

\noindent \textbf{Acknowledgments}

\noindent We thank N. Ofek and Y. Liu for valuable suggestions on
FPGA programming. CLZ is supported by the National Key Research and
Development Program of China (Grant No.2016YFA0301300, 2017YFA0304504)
and Anhui Initiative in Quantum Information Technologies (AHY130000).
LS acknowledges the support from National Natural Science Foundation
of China Grant No.11474177, National Key Research and Development
Program of China No.2017YFA0304303, and the Thousand Youth Fellowship
program in China. LS also thanks R. Vijay and his group for help on
the parametric amplifier measurements.

\clearpage{}

\newpage{}

\newpage{}
\end{document}


\onecolumngrid \global\long\def\thefigure{S\arabic{figure}}
 \setcounter{figure}{0} \global\long\def\thepage{S\arabic{page}}
 \setcounter{page}{1} \global\long\def\theequation{S.\arabic{equation}}
 \setcounter{equation}{0} 
\setcounter{section}{0}

\title{Supplementary Information ``Experimental repetitive quantum channel
simulation\char`\"{}}


\author{L.~Hu}

\thanks{These two authors contributed equally to this work.}

\affiliation{Center for Quantum Information, Institute for Interdisciplinary Information
Sciences, Tsinghua University, Beijing 100084, China}

\author{X.~Mu}

\thanks{These two authors contributed equally to this work.}

\affiliation{Center for Quantum Information, Institute for Interdisciplinary Information
Sciences, Tsinghua University, Beijing 100084, China}

\author{W.~Cai}

\affiliation{Center for Quantum Information, Institute for Interdisciplinary Information
Sciences, Tsinghua University, Beijing 100084, China}

\author{Y.~Ma}

\affiliation{Center for Quantum Information, Institute for Interdisciplinary Information
Sciences, Tsinghua University, Beijing 100084, China}

\author{Y.~Xu}

\affiliation{Center for Quantum Information, Institute for Interdisciplinary Information
Sciences, Tsinghua University, Beijing 100084, China}

\author{H.~Wang}

\affiliation{Center for Quantum Information, Institute for Interdisciplinary Information
Sciences, Tsinghua University, Beijing 100084, China}

\author{Y.~P.~Song}

\author{C.-L.~Zou}
\email{clzou321@ustc.edu.cn}


\affiliation{Key Laboratory of Quantum Information, CAS, University of Science
and Technology of China, Hefei, Anhui 230026, P. R. China}

\author{L.~Sun}
\email{luyansun@tsinghua.edu.cn}


\affiliation{Center for Quantum Information, Institute for Interdisciplinary Information
Sciences, Tsinghua University, Beijing 100084, China}

\maketitle

\section{Experimental System and Setup}

Our experimental device is measured in a dilution refrigerator with
a base temperature of about 10~mK. The device consists of an ancillary
transmon qubit dispersively coupled to both a readout cavity and a
storage cavity. The ancilla qubit is fabricated on a $c$-plane sapphire
(Al$_{2}$O$_{3}$) substrate with the standard double-angle evaporation
of aluminum and electron-beam lithography. The rectangular cavities
are made of high purity 5N5 aluminum, chemically etched for a better
coherence time~\cite{Reagor2013,Reagor2016PRB}. The ancillary transmon
qubit has a frequency $\omega_{\mathrm{q}}/2\pi=5.692$~GHz, an energy
relaxation time $T_{1}=30~{\mathrm{\mu}}$s, and a pure phasing time $T_{\phi}=120~{\mathrm{\mu}}$s.
The storage cavity has a frequency $\omega_{\mathrm{s}}/2\pi=7.634$~GHz,
a single-photon lifetime $T_{1}^{\mathrm{s}}=143~{\mathrm{\mu}}$s, and a
coherence time $T_{2}^{\mathrm{s}}=250~{\mathrm{\mu}}$s. Fock states $\ket{0}$
and $\ket{1}$ of the storage cavity constitute the two bases of a
photonic qubit on which channel simulations are realized. The dephasing,
damping, and arbitrary channels all rely on the dispersive interaction
$\chi_{\mathrm{s}}/2\pi=1.90$~MHz between the ancilla and the photonic
qubit. The readout cavity is at a frequency of $\omega_{\mathrm{r}}/2\pi=8.610$~GHz,
and has a lifetime of 44~ns and a dispersive interaction $\chi_{\mathrm{r}}/2\pi=3.65$~MHz
with the ancilla. With the help of a Josephson parametric Amplifier
(JPA)~\cite{Hatridge,Roy2015,Kamal,Murch} as the first stage of
amplification, quantum non-demolition single-shot measurements of
the ancilla qubit can be realized with high fidelities: $>99.9\%$
for the ground state $\ket{g}$ and $98.9\%$ for the excited state
$\ket{e}$ in a duration of 320~ns. Therefore, each readout measurement
throughout our experiment returns a digitized value of the ancilla
qubit state.

Fast real-time adaptive control is vital for the demonstrated channel
simulations. This is achieved through three field programmable gate
arrays (FPGA) with home-made logics, which are able to integrate readout
signal sampling, ancilla state estimation, and manipulation signal
generation together, and also allow us to individually control the
ancilla qubit, the photonic qubit, and the readout cavity. The latency
time, defined as the time interval between sending out the last point
of the readout signal and sending out the first point of the control
signal, is 340~ns (about $1\%$ of the ancilla qubit lifetime). This
time includes the signal travel time through the whole experimental
circuitry. More details of the experimental setup and FPGAs can be found
in our earlier report in Ref.~\onlinecite{Hu2018}.

\section{General Theory }

\subsection{Quantum channel and process matrix}

The Kraus representation of a quantum channel or quantum operation
can be written as~\cite{Nielsen}
\begin{equation}
\rho\mapsto\mathcal{E}\left(\rho\right)=\sum_{k=1}^{M}E_{k}\rho E_{k}^{\dagger},\label{eq:SI-channel}
\end{equation}
with $1\le M\le d^{2}$ ($d$ is the Hilbert space dimension) and
the Kraus operators satisfy
\begin{equation}
\sum_{k}E_{k}^{\dagger}E_{k}=\mathbf{I},
\end{equation}
where $\mathbf{I}$ is the unity matrix.

Representing the Kraus operators in a certain basis, the channel can
also be expressed based on the process $\chi$ matrix~\cite{Bhandari2016}
\begin{equation}
\rho\mapsto\sum_{m,n}\chi_{mn}\widetilde{E_{m}}\rho\widetilde{E_{n}}^{\dagger},\label{eq:SI-process}
\end{equation}
where the $\chi$ matrix contains $d^{4}-d^{2}$ independent parameters
under the constraints that
\begin{equation}
\sum_{m,n}\chi_{mn}\widetilde{E_{n}}^{\dagger}\widetilde{E_{m}}=\mathbf{I}.
\end{equation}
Usually, we choose Pauli matrices as the operation elements, with
the definition consistent with those in~\cite{Nielsen}:
\begin{equation}
\widetilde{E_{m}}\in\left\{ \mathbf{I},\mathbf{X},-i\mathbf{Y},\mathbf{Z}\right\}
\end{equation}
with
\begin{equation}
\mathbf{X}=\left[\begin{array}{cc}
0 & 1\\
1 & 0
\end{array}\right],\,\mathbf{Y}=\left[\begin{array}{cc}
0 & -i\\
i & 0
\end{array}\right],\,\mathbf{Z}=\left[\begin{array}{cc}
1 & 0\\
0 & -1
\end{array}\right],
\end{equation}
and the quantum state $\alpha\left|0\right\rangle +\beta\left|1\right\rangle $
is written in a vector notation as
\begin{equation}
\left[\begin{array}{c}
\alpha\\
\beta
\end{array}\right].
\end{equation}

In the following, we provide details of the basic channels studied
in this work.

\subsubsection{Identity channel}

The identity channel is trivial:
\begin{equation}
\mathcal{E}\left(\rho\right)=\rho,
\end{equation}
and the corresponding process matrix is
\begin{equation}
\chi_{\mathrm{I}}=\left[\begin{array}{cccc}
1 & 0 & 0 & 0\\
0 & 0 & 0 & 0\\
0 & 0 & 0 & 0\\
0 & 0 & 0 & 0
\end{array}\right].
\end{equation}
To characterize the ability of quantum channel $\chi$ for keeping
quantum information, we introduce the process fidelity as
\begin{equation}
F_{\chi}=\mathrm{Tr}\left(\chi_{\mathrm{I}}\chi\right)
\end{equation}

\subsubsection{Depolarization channel}

The depolarization channel is defined as~\cite{Nielsen}
\begin{equation}
\mathcal{E}\left(\rho\right)=\left(1-p\right)\mathbf{I}\rho\mathbf{I}+\frac{p}{3}\left(\mathbf{X}\rho\mathbf{X}+\mathbf{Y}\rho\mathbf{Y}+\mathbf{Z}\rho\mathbf{Z}\right),
\end{equation}
and the corresponding process matrix is
\begin{equation}
\chi_{\mathrm{dpl}}=\left[\begin{array}{cccc}
1-p & 0 & 0 & 0\\
0 & \frac{p}{3} & 0 & 0\\
0 & 0 & \frac{p}{3} & 0\\
0 & 0 & 0 & \frac{p}{3}
\end{array}\right].
\end{equation}
The effect of this channel is to mix the input
state with the completely mixed state $\mathbf{I}/2$.

\subsubsection{Qubit dephasing channel}

The dephasing channel is defined as~\cite{Nielsen}
\begin{equation}
\mathcal{E}\left(\rho\right)=E_{0}\rho E_{0}^{\dagger}+E_{1}\rho E_{1}^{\dagger}\label{eq:SI-dephasingchannel}
\end{equation}
with
\begin{eqnarray}
E_{0}=\sqrt{p}\mathbf{I},\,E_{1}=\sqrt{1-p}\mathbf{Z},
\end{eqnarray}
where $p\in\left[0,1\right]$. According to Eq.$\,$(\ref{eq:SI-process}),
we can get the $\chi$ matrix for a dephasing channel as
\begin{equation}
\chi_{\mathrm{dph}}=\left[\begin{array}{cccc}
p & 0 & 0 & 0\\
0 & 0 & 0 & 0\\
0 & 0 & 0 & 0\\
0 & 0 & 0 & 1-p
\end{array}\right].
\end{equation}
After repetitively implementing the channel for $n$ times, we have
the process matrix for $\mathcal{E}^{n}$
\begin{equation}
\chi_{\mathrm{dph}}^{\left(n\right)}=\left[\begin{array}{cccc}
2^{n-1}(p-\frac{1}{2})^{n}+\frac{1}{2} & 0 & 0 & 0\\
0 & 0 & 0 & 0\\
0 & 0 & 0 & 0\\
0 & 0 & 0 & \frac{1}{2}-2^{n-1}(p-\frac{1}{2})^{n}
\end{array}\right].
\end{equation}
Therefore, the process fidelity decays exponentially with $n$ as
\begin{equation}
F_{\chi}=\frac{1}{2}\left[1+\left(2p-1\right)^{n}\right],
\end{equation}
which converges to $F_{\chi}=\frac{1}{2}$ for $n\gg1$.

\subsubsection{Qubit damping channel}

The amplitude damping channel is defined as~\cite{Nielsen}
\begin{equation}
\mathcal{E}\left(\rho\right)=E_{0}\rho E_{0}^{\dagger}+E_{1}\rho E_{1}^{\dagger},
\end{equation}
with
\begin{eqnarray}
E_{0}=\left[\begin{array}{cc}
1 & 0\\
0 & \sqrt{1-\gamma}
\end{array}\right],\,E_{1}=\left[\begin{matrix}0 & \sqrt{\gamma}\\
0 & 0
\end{matrix}\right],\label{eq:SI-dampingchannel}
\end{eqnarray}
where $\gamma\in\left[0,1\right]$. According to Eq.$\,$(\ref{eq:SI-process}),
we can get the $\chi$ matrix for a damping channel
\begin{equation}
\chi_{{\rm dmp}}=\left[\begin{array}{cccc}
\frac{\sqrt{1-\gamma}}{2}-\frac{\gamma}{4}+\frac{1}{2} & 0 & 0 & \frac{\gamma}{4}\\
0 & \frac{\gamma}{4} & -\frac{\gamma}{4} & 0\\
0 & -\frac{\gamma}{4} & \frac{\gamma}{4} & 0\\
\frac{\gamma}{4} & 0 & 0 & \frac{1}{2}-\frac{\sqrt{1-\gamma}}{2}-\frac{\gamma}{4}
\end{array}\right].
\end{equation}
After repetitively implementing the channel for $n$ times, we have
the process matrix for $\mathcal{E}^{n}$
\begin{equation}
\chi_{{\rm dmp}}^{\left(n\right)}=\left[\begin{array}{cccc}
\frac{[1+(1-\gamma)^{\frac{n}{2}}]^{2}}{4} & 0 & 0 & \frac{[1-(1-\gamma)^{n}]}{4}\\
0 & \frac{[1-(1-\gamma)^{n}]}{4} & -\frac{[1-(1-\gamma)^{n}]}{4} & 0\\
0 & -\frac{[1-(1-\gamma)^{n}]}{4} & \frac{[1-(1-\gamma)^{n}]}{4} & 0\\
\frac{[1-(1-\gamma)^{n}]}{4} & 0 & 0 & \frac{[1-(1-\gamma)^{\frac{n}{2}}]^{2}}{4}
\end{array}\right].
\end{equation}
Therefore, the process fidelity decays bi-exponentially with $n$
as
\begin{equation}
F_{\chi}=\frac{1}{4}+\frac{1}{2}(1-\gamma)^{\frac{n}{2}}+\frac{1}{4}(1-\gamma)^{n},
\end{equation}
which converges to $F_{\chi}=\frac{1}{4}$ for $n\gg1$.

\subsection{Correspondence between quantum channel simulation and master equation }

\subsubsection{Liouvillian and repetitive quantum channel simulation}

Here, we explain the correspondence between digital quantum channel
simulation and master equation for continuous open system dynamics.
We start the discussion with a simple example, in which the system
couples to a Markovian environment through the operator $o$ as
\begin{equation}
H=H_{0}+\sum_{\omega}\omega b_{\omega}^{\dagger}b_{\omega}+\sum_{\omega}\left(g_{\omega}o^{\dagger}b_{\omega}+h.c.\right),
\end{equation}
where the bosonic operator $b_{\omega}$ denotes the harmonic oscillator
mode in the bath with frequency of $\omega$, and $g_{\omega}$ is
the corresponding coupling strength. Applying the Born approximation,
we obtain the evolution of the system density matrix as \cite{Carmichael2003}
\begin{eqnarray}
\frac{d}{dt}\rho_{\mathrm{s}} & = & -i\left[H_{0},\rho_{\mathrm{s}}\right]\nonumber \\
 &  & -\left\{ \int_{0}^{t}d\tau\left[oo^{\dagger}\rho_{\mathrm{s}}\left(t-\tau\right)-o^{\dagger}\rho_{\mathrm{s}}\left(t-\tau\right)o\right]\sum_{\omega}\left|g_{\omega}^{2}\right|\left\langle b_{\omega}^{\dagger}\left(t\right)b_{\omega}\left(t-\tau\right)\right\rangle +h.c.\right\} \nonumber \\
 &  & -\left\{ \int_{0}^{t}d\tau\left[o^{\dagger}o\rho_{\mathrm{s}}\left(t-\tau\right)-o\rho_{\mathrm{s}}\left(t-\tau\right)o^{\dagger}\right]\sum_{\omega}\left|g_{\omega}^{2}\right|\left\langle b_{\omega}\left(t\right)b_{\omega}^{\dagger}\left(t-\tau\right)\right\rangle +h.c.\right\} .
\end{eqnarray}
For a low-temperature reservoir, $\left\langle b_{\omega}^{\dagger}\left(t\right)b_{\omega}\left(t-\tau\right)\right\rangle \approx0$
and $\left\langle b_{\omega}\left(t\right)b_{\omega}^{\dagger}\left(t-\tau\right)\right\rangle =\delta\left(\tau\right)$. If we define $\kappa=\frac{1}{2}\int d\omega\xi\left(\omega\right)\left|g_{\omega}^{2}\right|$
with $\xi\left(\omega\right)$ being the density of states in the bath,
we obtain the master equation in the Lindblad form as~\cite{Petruccione2002,gardiner2004quantum}
\begin{align}
\frac{d}{dt}\rho & =-i\left[H_{0},\rho\right]+\kappa\left(2o\rho o^{\dagger}-o^{\dagger}o\rho-\rho o^{\dagger}o\right)\label{eq:SI-masterEq}\\
 & =\mathcal{L}_{0}\left(\rho\right)+\mathcal{L}_{1}\left(\rho\right),
\end{align}
with Liouvillian $\mathcal{L}_{0}$ and $\mathcal{L}_{1}$ describing
the coherent and incoherent evolutions, respectively. For $dt\rightarrow0$,
we have
\begin{align}
\rho\left(t+dt\right)-\rho\left(t\right) & =\mathcal{L}_{0}\left(\rho\right)dt+\mathcal{L}_{1}\left(\rho\right)dt+\mathcal{O}\left[\left(H_{0}dt\right)^{2}\rho\right]+\mathcal{O}\left[\left(\kappa dt\right)^{2}\rho\right].
\end{align}

Through channel simulation, we can implement unitary channel
\begin{align}
\mathcal{E}_{0}\left(\rho\right) & =e^{-iHdt}\rho e^{iHdt}\\
 & =\rho-i\left[H_{0}dt,\rho\right]+\mathcal{O}\left[\left(H_{0}dt\right)^{2}\rho\right]\\
 & =\rho+\mathcal{L}_{0}\left(\rho\right)dt+\mathcal{O}\left[\left(H_{0}dt\right)^{2}\rho\right]
\end{align}
through arbitrary unitary control \cite{Krastanov2015}, and non-unitary
channel
\begin{align}
\mathcal{E}_{1}\left(\rho\right) & =E_{1,0}\rho E_{1,0}^{\dagger}+E_{1,1}\rho E_{1,1}^{\dagger}\\
 & =\rho+\kappa dt\left(2o\rho o^{\dagger}-o^{\dagger}o\rho-\rho o^{\dagger}o\right)+\mathcal{O}\left[\left(\kappa dt\right)^{2}\rho\right]\\
 & =\mathcal{L}_{1}\left(\rho\right)dt+\mathcal{O}\left[\left(\kappa dt\right)^{2}\rho\right]
\end{align}
with
\begin{align}
E_{1,0} & =\sqrt{2\kappa dt}o,\\
E_{1,1} & =\mathbf{I}-\kappa dto^{\dagger}o.
\end{align}
Therefore, by sequentially implementing these two channels
\begin{align}
\mathcal{E}_{1}\circ\mathcal{E}_{0}\left(\rho\right) & =\mathcal{L}_{0}\left(\rho\right)dt+\mathcal{L}_{1}\left(\rho\right)dt+\mathcal{O}\left[\left(H_{0}dt\right)^{2}\rho\right]+\mathcal{O}\left[\left(\kappa dt\right)^{2}\rho\right]\\
 & \approx e^{dt\left(\mathcal{L}_{0}+\mathcal{L}_{1}\right)}\rho,
\end{align}
we obtain the effective implementation of the Liouvillian with an
error to the order of $\left(H_{0}dt\right)^{2}+\left(\kappa dt\right)^{2}$.

The above derivations can be easily generalized to multiple-Liouvillian
case, i.e. $\sum_{j}\mathcal{L}_{j}$. As proved in Refs.~\cite{Kliesch2011,Sweke2014},
the effective continuous open quantum system dynamics can be realized
by repetitively implementing the quantum channels which correspond
to each individual Liouvillian alternatively.

\subsubsection{Qubit dephasing channel}

When $o=\mathbf{Z}$ and $H=0$ in Eq.$\,$(\ref{eq:SI-masterEq}),
the master equation corresponds to a simple Markovian dephasing channel
on a qubit. According to the analysis above, when
\begin{align}
E_{1,0} & =\sqrt{2\kappa dt}\mathbf{Z},\\
E_{1,1} & =\sqrt{1-2\kappa dt}\mathbf{I},
\end{align}
the repetitive implementation of the channel (in the limit of $\kappa dt\ll1$)
will give us an effective continuous dephasing process. According
to the definition of a dephasing channel (Eq.$\,$(\ref{eq:SI-dephasingchannel})),
it requires that $p=1-2\kappa dt$. It also indicates that the
implementation of the digital dephasing channel simulation with $1-p\ll1$
and a duration of $\tau_{0}$ leads to an effective continuous dephasing
Liouvillian with a dephasing rate $\kappa=\left(1-p\right)/2\tau_{0}$,
and the corresponding phase coherence time is $T_{2}=1/4\kappa$.

\subsubsection{Qubit damping channel}

Similarly, when
\begin{equation}
o=\sigma_{-}=\left[\begin{array}{cc}
0 & 1\\
0 & 0
\end{array}\right]
\end{equation}
and $H=0$ in Eq.$\,$(\ref{eq:SI-masterEq}), the master equation
corresponds to a simple Markovian excited state damping channel
on a qubit. If
\begin{align}
E_{1,0} & =\sqrt{2\kappa dt}\sigma_{-},\\
E_{1,1} & =\mathbf{I}-\kappa dt\frac{\mathbf{I}-\mathbf{Z}}{2},
\end{align}
the repetitive implementation of the channel (in the limit of $\kappa dt\ll1$)
will give us an effective continuous damping process. According to
the definition of a damping channel Eq.$\,$(\ref{eq:SI-dampingchannel}),
it requires that $\gamma=2\kappa dt$. It also indicates that the
implementation of the digital damping channel simulation with $\gamma\ll1$
and a duration of $\tau_{0}$ leads to an effective continuous damping
Liouvillian with a damping rate $\kappa=\gamma/2\tau_{0}$.

\section{Experimental Quantum Channel Simulations}

\subsection{The dephasing channel simulation}

\begin{figure*}[hbt]
\centering \includegraphics{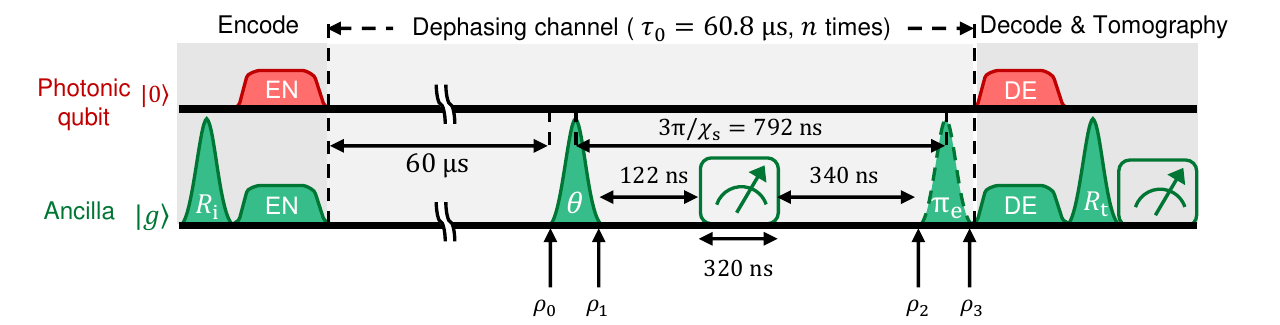} \caption{\textbf{Experimental sequence for a dephasing channel simulation.}
Density matrices at the arrows are shown in detail in the text.}
\label{fig:dephasing_exp}
\end{figure*}

In our experiment, the dephasing channel is realized by using the
control pulse sequence in Fig.$\,$2(a) in the main text. To explain
the protocol, we analyze the procedures in detail as follows.
\begin{enumerate}
\item The ancilla qubit is initialized to the ground state $\rho_{\mathrm{a}}=\left|g\right\rangle \left\langle g\right|$,
and the photonic qubit is encoded in $\rho_{\mathrm{in}}$. Then the
input of the system is
\begin{equation}
\rho_{0}=\rho_{\mathrm{a}}\otimes\rho_{{\rm in}}=\left[\begin{array}{cc}
1 & 0\\
0 & 0
\end{array}\right]\otimes\left[\begin{array}{cc}
a & b\\
c & d
\end{array}\right].
\end{equation}
\item After applying $X_{\theta}$ (a unitary operation $U_{X}\left(\theta\right)=e^{-i\frac{\theta}{2}\mathbf{X}}$)
to the ancilla qubit, we have
\begin{align}
\rho_{0}\mapsto\rho_{1} & =U_{\mathbf{X}}\left(\theta\right)\rho_{0}U_{\mathbf{X}}^{\dagger}\left(\theta\right)\\
 & =\left[\begin{array}{cc}
\cos^{2}\left(\frac{\theta}{2}\right) & i\sin\left(\frac{\theta}{2}\right)\cos\left(\frac{\theta}{2}\right)\\
-i\sin\left(\frac{\theta}{2}\right)\cos\left(\frac{\theta}{2}\right) & \sin^{2}\left(\frac{\theta}{2}\right)
\end{array}\right]\otimes\left[\begin{array}{cc}
a & b\\
c & d
\end{array}\right].
\end{align}
\item According to the Hamiltonian $H=\omega_{s}a^{\dagger}a+\omega_{a}|e\rangle\langle e|-\chi_{{\rm {s}}}a^{\dagger}a|e\rangle\langle e|$,
the $\ket{e}\ket{1}$ state gains an extra phase $e^{i\chi_{{\rm {s}}}t}$
for a duration of $t$. Thus, waiting for a duration of $t=3\pi/\chi_{{\rm {s}}}$,
a controlled-Z (CZ) gate between the ancilla and the photonic qubit
can be realized as
\begin{align}
\rho_{1}\mapsto\rho_{2} & =U_{\mathrm{CZ}}\rho_{1}U_{\mathrm{CZ}}^{\dagger}\\
 & =\left[\begin{array}{cc}
\cos^{2}\left(\frac{\theta}{2}\right)\left[\begin{array}{cc}
a & b\\
c & d
\end{array}\right] & i\sin\left(\frac{\theta}{2}\right)\cos\left(\frac{\theta}{2}\right)\left[\begin{array}{cc}
a & -b\\
c & -d
\end{array}\right]\\
-i\sin\left(\frac{\theta}{2}\right)\cos\left(\frac{\theta}{2}\right)\left[\begin{array}{cc}
a & b\\
-c & -d
\end{array}\right] & \sin^{2}\left(\frac{\theta}{2}\right)\left[\begin{array}{cc}
a & -b\\
-c & d
\end{array}\right]
\end{array}\right].
\end{align}
\item When tracing out the ancilla qubit state, we have the output
state of the photonic qubit as
\begin{align}
\rho_{\mathrm{out}} & =\mathrm{Tr}_{\mathrm{a}}\left(\rho_{2}\right)\\
 & =\cos^{2}\left(\frac{\theta}{2}\right)\left[\begin{array}{cc}
a & b\\
c & d
\end{array}\right]+\sin^{2}\left(\frac{\theta}{2}\right)\left[\begin{array}{cc}
a & -b\\
-c & d
\end{array}\right]\\
 & =\left[\begin{array}{cc}
a & b{\rm cos}\left(\theta\right)\\
c{\rm cos}\left(\theta\right) & d
\end{array}\right].
\end{align}
\item To discard the ancilla qubit state, the ancilla
is measured after the CZ gate and a conditional $X$ operation is
implemented when the measurement outcome is $\ket{e}$. This measurement
and adaptive operation on the ancilla qubit can be described as
\begin{equation}
\mathcal{E}_{\mathrm{MF}}\left(\rho_{\mathrm{a}}\right):\,\rho_{\mathrm{a}}\mapsto\left|g\right\rangle \left\langle g\right|\rho_{\mathrm{a}}\left|g\right\rangle \left\langle g\right|+X\left|e\right\rangle \left\langle e\right|\rho_{\mathrm{a}}\left|e\right\rangle \left\langle e\right|X^{\dagger},
\end{equation}
i.e.
\begin{equation}
\mathcal{E}_{\mathrm{MF}}\left(\rho_{\mathrm{a}}\right):\,\rho_{\mathrm{a}}\mapsto\left[\begin{array}{cc}
1 & 0\\
0 & 0
\end{array}\right]\mathrm{Tr}_{\mathrm{a}}\left(\rho_{\mathrm{a}}\right).
\end{equation}
Thus, the system state before decoding is
\begin{align}
\rho_{2}\mapsto\rho_{3} & =\mathcal{E}_{\mathrm{MF}}\left(\rho_{2}\right)=\left[\begin{array}{cc}
1 & 0\\
0 & 0
\end{array}\right]\otimes\mathrm{Tr}_{\mathrm{a}}\left(\rho_{2}\right).\\
 & =\left[\begin{array}{cc}
1 & 0\\
0 & 0
\end{array}\right]\otimes\left[\begin{array}{cc}
a & b{\rm cos}\left(\theta\right)\\
c{\rm cos}\left(\theta\right) & d
\end{array}\right].
\end{align}
\end{enumerate}
Comparing the input and output of the experimental procedure, we find
the obtained channel is
\begin{equation}
\mathcal{E}_{\mathrm{anc}\otimes\mathrm{target}}=\mathcal{I}_{\mathrm{anc}}\otimes\mathcal{E}_{\mathrm{target}},
\end{equation}
where $\mathcal{I}:\,\rho\mapsto\rho$ is the identity channel
and
\begin{equation}
\mathcal{E}_{\mathrm{target}}:\,\rho\mapsto\cos^{2}\left(\frac{\theta}{2}\right)\mathbf{I}\rho\mathbf{I}^{\dagger}+\sin^{2}\left(\frac{\theta}{2}\right)\mathbf{Z}\rho\mathbf{Z}^{\dagger}.
\end{equation}
According to the definition in last section {[}Eq.$\,$(\ref{eq:SI-dephasingchannel}){]},
the dephasing channel with $p=\cos^{2}\left(\frac{\theta}{2}\right)$
is realized.

The effects of the repetitive dephasing channel simulation and the intrinsic
decoherence can simply be added together, which leads to an effective
channel coherence time
\begin{equation}
\frac{1}{T_{2}^{c}}=-\frac{{\rm ln}\left(2\cos^{2}\frac{\theta}{2}-1\right)}{\tau_{0}}+\frac{1}{T_{2}^{0}},
\end{equation}
where $T_{2}^{0}$ is the intrinsic coherence time (when $\theta=0^{\circ}$)
and $\tau_{0}$ is the time interval. In the main text, we get $T_{2}^{0}=240\pm4~{\mathrm{\mu}}$s,
which agrees well with $T_{2}^{{\rm {s}}}=250~{\mathrm{\mu}}$s.

In our experiments, the duration of the digital quantum channel simulation
is variable. When $\tau_{0}\ll T_{1}^{0},T_{2}^{0}$, according to
the discussions on the correspondence between continuous Liouvillian
and repetitive channel simulation, the intrinsic dephasing and damping
of the photonic qubit can also be treated as two channels. Therefore,
the resulting effective dynamics when $\theta\ll1$ can be approximated
by a simple combination of the decoherence processes:
\begin{align}
\frac{1}{T_{2}^{c}} & \approx\frac{2\sin^{2}\left(\frac{\theta}{2}\right)}{\tau_{0}}+\frac{1}{T_{2}^{0}}.
\end{align}

\subsection{The damping channel simulation}

\begin{figure*}[hbt]
\centering \includegraphics{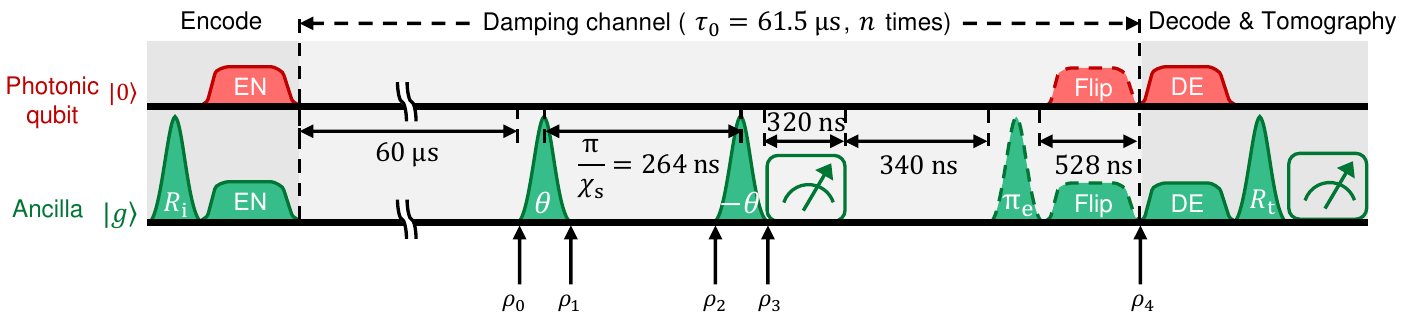} \caption{\textbf{Experimental sequence for a damping channel simulation.} Density
matrices at the arrows are shown in detail in the text.}
\label{fig:damping_exp}
\end{figure*}

In our experiment, the damping channel is realized by using the control
pulse sequence in Fig.$\,$3(a) in the main text. To explain the protocol,
we analyze the procedures in detail as follows.
\begin{enumerate}
\item The initial state of photonic qubit is $\rho_{\mathrm{in}}$, and the
ancilla qubit is initialized to ground state $\rho_{\mathrm{a}}=\left|g\right\rangle \left\langle g\right|$.
So, the input system of the system is
\begin{equation}
\rho_{0}=\rho_{\mathrm{a}}\otimes\rho_{{\rm in}}=\left[\begin{array}{cc}
1 & 0\\
0 & 0
\end{array}\right]\otimes\left[\begin{array}{cc}
a & b\\
c & d
\end{array}\right].
\end{equation}
\item After applying $X_{\theta}$ ($U_{X}\left(\theta\right)=e^{-i\frac{\theta}{2}\mathbf{X}}$)
to the ancilla qubit, we have
\begin{align}
\rho_{0}\mapsto\rho_{1} & =U_{\mathbf{X}}\left(\theta\right)\rho_{0}U_{\mathbf{X}}^{\dagger}\left(\theta\right)\\
 & =\left[\begin{array}{cc}
\cos^{2}\left(\frac{\theta}{2}\right) & i\sin\left(\frac{\theta}{2}\right)\cos\left(\frac{\theta}{2}\right)\\
-i\sin\left(\frac{\theta}{2}\right)\cos\left(\frac{\theta}{2}\right) & \sin^{2}\left(\frac{\theta}{2}\right)
\end{array}\right]\otimes\left[\begin{array}{cc}
a & b\\
c & d
\end{array}\right].
\end{align}
\item Due to the dispersive ancilla-photonic qubit interaction, a controlled-Z
(CZ) gate can be obtained after a duration of $\pi/\chi_{{\rm {s}}}$,
thus
\begin{align}
\rho_{1}\mapsto\rho_{2} & =U_{\mathrm{CZ}}\rho_{1}U_{\mathrm{CZ}}^{\dagger}\\
 & =\left[\begin{array}{cc}
\cos^{2}\left(\frac{\theta}{2}\right)\left[\begin{array}{cc}
a & b\\
c & d
\end{array}\right] & i\sin\left(\frac{\theta}{2}\right)\cos\left(\frac{\theta}{2}\right)\left[\begin{array}{cc}
a & -b\\
c & -d
\end{array}\right]\\
-i\sin\left(\frac{\theta}{2}\right)\cos\left(\frac{\theta}{2}\right)\left[\begin{array}{cc}
a & b\\
-c & -d
\end{array}\right] & \sin^{2}\left(\frac{\theta}{2}\right)\left[\begin{array}{cc}
a & -b\\
-c & d
\end{array}\right]
\end{array}\right].
\end{align}
\item After applying the second $X_{-\theta}$ (a unitary operation $U_{X}\left(-\theta\right)=e^{i\frac{\theta}{2}\mathbf{X}}$)
to the ancilla qubit, we have
\begin{align}
\rho_{2}\mapsto\rho_{3} & =U_{\mathbf{X}}\left(-\theta\right)\rho_{2}U_{\mathbf{X}}^{\dagger}\left(-\theta\right)\\
 & =\left[\begin{array}{cc}
\left[\begin{array}{cc}
a & b{\rm cos}\left(\theta\right)\\
c{\rm cos}\left(\theta\right) & d{\rm cos}^{2}\left(\theta\right)
\end{array}\right] & \left[\begin{array}{cc}
0 & -ib{\rm sin}\left(\theta\right)\\
0 & -id{\rm cos}\left(\theta\right){\rm sin}\left(\theta\right)
\end{array}\right]\\
\left[\begin{array}{cc}
0 & 0\\
ic{\rm sin}\left(\theta\right) & id{\rm cos}\left(\theta\right){\rm sin}\left(\theta\right)
\end{array}\right] & \left[\begin{array}{cc}
0 & 0\\
0 & d{\rm sin}^{2}\left(\theta\right)
\end{array}\right]
\end{array}\right].
\end{align}
\item After implementing a measurement-based adaptive operation on the photonic
qubit, we have
\begin{align}
\rho_{3}\mapsto\rho_{4} & =\left[\left|g\right\rangle \left\langle g\right|\otimes\mathbf{I}\right]\rho_{3}\left[\left|g\right\rangle \left\langle g\right|\otimes\mathbf{I}\right]^{\dagger}+\left[\left|e\right\rangle \left\langle e\right|\otimes\mathbf{X}\right]\rho_{3}\left[\left|e\right\rangle \left\langle e\right|\otimes\mathbf{X}\right]^{\dagger}\\
 & =\left[\begin{array}{cc}
a & b{\rm cos}\left(\theta\right)\\
c{\rm cos}\left(\theta\right) & d{\rm cos}^{2}\left(\theta\right)
\end{array}\right]+\left[\begin{array}{cc}
d{\rm sin}^{2}\left(\theta\right) & 0\\
0 & 0
\end{array}\right]\\
 & =\left[\begin{array}{cc}
a+d{\rm sin}^{2}\left(\theta\right) & b{\rm cos}\left(\theta\right)\\
c{\rm cos}\left(\theta\right) & d{\rm cos}^{2}\left(\theta\right)
\end{array}\right].
\end{align}
\end{enumerate}
Comparing to the definition for a damping channel (Eq.$\,$\ref{eq:SI-dampingchannel}),
we realize a damping channel to the photonic qubit with $\gamma=\sin^{2}\left(\theta\right)$.

Similar to the case of the dephasing channel, we have
\begin{eqnarray}
\frac{1}{T_{1}^{c}} & = & -\frac{{\rm ln}({\rm cos}^{2}\theta)}{\tau_{0}}+\frac{1}{T_{1}^{0}},\\
\frac{1}{T_{2}^{c}} & = & -\frac{{\rm ln}({\rm cos}^{2}\theta)}{2\tau_{0}}+\frac{1}{T_{2}^{0}},
\end{eqnarray}
where $\tau_{0}$ is the time interval, and $T_{1}^{0}$, $T_{2}^{0}$
are the intrinsic decoherence times of the photonic qubit ($\theta=0^{\circ}$).
In the main text, we get $T_{1}^{0}=137\pm4~{\mathrm{\mu}}$s and $T_{2}^{0}=237\pm3~{\mathrm{\mu}}$s,
which agree well with $T_{1}^{{\rm {s}}}=143~{\mathrm{\mu}}$s and $T_{2}^{{\rm {s}}}=250~{\mathrm{\mu}}$s,
respectively.

\subsection{Arbitrary channel simulation}

\subsubsection{Simulation of arbitrary quantum channel}

Following the proposal by Wang et al.~\cite{Wang2013}, we realize
the arbitrary single-qubit channel by a convex combination of quasiextreme
channels, where only a single ancilla qubit is required. As shown
in Fig.$\,$1(c) in the main text, the ancilla qubit is firstly prepared
to a superposition state as
\begin{equation}
\rho_{\mathrm{a}}=X\left(\gamma_{p}\right)\left|g\right\rangle \left\langle g\right|X^{\dagger}\left(\gamma_{p}\right),
\end{equation}
where $X\left(\theta\right)=e^{-i\frac{\theta}{2}\mathbf{X}}=\cos\frac{\theta}{2}\mathbf{I}-i\sin\frac{\theta}{2}\mathbf{X}$.
Then, a projective measurement on the ancilla qubit in the computational
basis is applied, and different channel branches are implemented according
to the measurement result. This can be represented by the quantum
instrument~\cite{Wilde2013}
\begin{equation}
\mathcal{E}:\,\rho\mapsto\mathcal{E}_{1}\otimes\left|g\right\rangle \left\langle g\right|+\mathcal{E}_{2}\otimes\left|e\right\rangle \left\langle e\right|.
\end{equation}
Here, $\mathcal{E}_{j}$ with $j=0,1$ is the quasiextreme channel,
which can be represented as
\begin{eqnarray}
\mathcal{E}_{j}:\,\rho & \mapsto & U\left(\overrightarrow{\varphi}\right)\left[\left|g\right\rangle \left\langle g\right|\otimes I\right]X\left(2\gamma_{2}\right)U_{CZ}\left[X\left(2\gamma_{1}\right)\otimes U\left(\overrightarrow{\delta}\right)\right]\rho\left\{ U\left(\overrightarrow{\varphi}\right)\left[\left|g\right\rangle \left\langle g\right|\otimes I\right]X\left(2\gamma_{2}\right)U_{CZ}\left[U\left(\overrightarrow{\delta}\right)\otimes X\left(2\gamma_{1}\right)\right]\right\} ^{\dagger}\nonumber \\
 &  & +U\left(\overrightarrow{\varphi}\right)\left[\left|e\right\rangle \left\langle e\right|\otimes X\right]X\left(2\gamma_{2}\right)U_{CZ}\left[X\left(2\gamma_{1}\right)\otimes U\left(\overrightarrow{\delta}\right)\right]\rho\left\{ U\left(\overrightarrow{\varphi}\right)\left[\left|e\right\rangle \left\langle e\right|\otimes X\right]X\left(2\gamma_{2}\right)U_{CZ}\left[U\left(\overrightarrow{\delta}\right)\otimes X\left(2\gamma_{1}\right)\right]\right\} ^{\dagger}
\end{eqnarray}
with parameters $\left\{ \overrightarrow{\delta^{\left(j\right)}},\,\gamma_{1}^{\left(j\right)},\,\gamma_{2}^{\left(j\right)},\,\overrightarrow{\varphi^{\left(j\right)}}\right\} $,
and unitary operation
\begin{equation}
U\left(\overrightarrow{\varphi}\right)=\left[\begin{array}{cc}
\cos\frac{\varphi_{1}}{2}-i\sin\frac{\varphi_{1}}{2}\cos\frac{\varphi_{2}}{2} & -i\sin\frac{\varphi_{1}}{2}\sin\frac{\varphi_{2}}{2}e^{-i\varphi_{3}}\\
-i\sin\frac{\varphi_{1}}{2}\sin\frac{\varphi_{2}}{2}e^{i\varphi_{3}} & \cos\frac{\varphi_{1}}{2}+i\sin\frac{\varphi_{1}}{2}\cos\frac{\varphi_{2}}{2}
\end{array}\right]
\end{equation}
with $\varphi_{1,2,3}\in\left[0,2\pi\right]$. Therefore, there are
17 parameters in total, including $\gamma_{p}$ and two sets of $8$
parameters for quasiextreme channels. Actually, the dephasing and
damping channels are two examples of quasiextreme channels. For a
dephasing channel, we have $\gamma_{p}=0$ and $\left\{ \overrightarrow{\delta^{\left(0\right)}},\,\gamma_{1}^{\left(0\right)},\,\gamma_{2}^{\left(0\right)},\,\overrightarrow{\varphi^{\left(0\right)}}\right\} =\left\{ \left(0,0,0\right),0,\theta,\left(0,0,1\right)\right\} $.
For a damping channel, we have $\gamma_{p}=0$ and $\left\{ \overrightarrow{\delta^{\left(0\right)}},\,\gamma_{1}^{\left(0\right)},\,\gamma_{2}^{\left(0\right)},\,\overrightarrow{\varphi^{\left(0\right)}}\right\} =\left\{ \left(0,0,0\right),\theta,-\theta,\left(0,0,1\right)\right\} $.

\subsubsection{Random qubit channel}

In our experiment, to demonstrate the ability of simulating arbitrary
single-qubit channel, we first generate a single-qubit channel numerically
with 12 random numbers in a target process matrix $\chi_{\mathrm{T}}$ that
satisfy certain constraints~\cite{Bhandari2016}. For the experimental
results shown in the main text, we generated 6 random qubit channels as
the targets for arbitrary quantum channel simulation. Here, we show the
target $\chi_{\mathrm{T}}$ for channel arbi.~1 as an example
\begin{equation}
\chi_{\mathrm{T}}^{\left(1\right)}=\left(\begin{array}{cccc}
0.362762\, & 0.0143269\,+0.0767813i & -0.0305766+0.075406i & -0.00589476-0.134834i\\
0.0143269\,-0.0767813i & 0.22118 & 0.00589476\,+0.0979485i & -0.0331165-0.075406i\\
-0.0305766-0.075406i & 0.00589476\,-0.0979485i & 0.300734 & 0.0143269\,+0.027254i\\
-0.00589476+0.134834i & -0.0331165+0.075406i & 0.0143269\,-0.027254i & 0.115325
\end{array}\right).
\end{equation}

For a given target process matrix $\chi_{\mathrm{T}}$, the $17$
parameters for the arbitrary channel simulation are determined through
numerical optimization that reduces the difference between $\chi_{\mathrm{sim}}$
and $\chi_{\mathrm{T}}$. Here, $\chi_{\mathrm{sim}}$ is the simulation
result for the quantum circuit shown in Fig.$\,$1(c) in the main text.
Then, we implement the channel with the experimental sequence shown in Fig.$\,$4(a)
in the main text with the optimized $17$ parameters. The obtained
experimental quantum channel $\chi_{\mathrm{E}}$ for the arbi.~1 is
\begin{equation}
\chi_{\mathrm{E}}^{\left(1\right)}=\left(\begin{array}{cccc}
0.352118 & 0.0205553\,+0.0697286i & -0.0117642+0.0543421i & -0.00241937-0.107611i\\
0.0205553\,-0.0697286i & 0.244587 & 0.00241937\,+0.0722139i & -0.0386366-0.0543421i\\
-0.0117642-0.0543421i & 0.00241937\,-0.0722139i & 0.251896 & 0.0205553\,+0.0262903i\\
-0.00241937+0.107611i & -0.0386366+0.0543421i & 0.0205553\,-0.0262903i & 0.151398
\end{array}\right).
\end{equation}

\subsubsection{Characterization of quantum channels}

To characterize the performance of the experiment for arbitrary
quantum channel simulation, we introduce two different figures of merit. The first one is the fidelity of state generation
\begin{equation}
F_{\mathrm{G}}=\mathrm{inf}_{\rho}\sqrt{\sqrt{\mathcal{E}_{\mathrm{T}}\left(\rho\right)}\left[\mathcal{E}_{\mathrm{E}}\left(\rho\right)\right]\sqrt{\mathcal{E}_{\mathrm{T}}\left(\rho\right)}}.
\end{equation}
The physical meaning of $F_{\mathrm{G}}$ is the fidelity of the output
quantum state $\mathcal{E}_{\mathrm{E}}\left(\rho\right)$ when we
use the quantum channel to manipulate the quantum state for a given
input compared with the result of the target quantum channel $\mathcal{E}_{\mathrm{T}}\left(\rho\right)$.
Therefore, $F_{\mathrm{G}}$ is the worst fidelity of the quantum state generated
by the experimental channel.

The other one is the measure of discriminating quantum channels.
For two channels $\mathcal{E}_{1}$ and $\mathcal{E}_{2}$, the diamond
distance is~\cite{Wilde2013}
\begin{equation}
D_{\diamond}=\left|\left|\mathcal{E}_{1}-\mathcal{E}_{2}\right|\right|_{\diamond}=\mathrm{sup}_{\rho}\left|\left|\left(\mathcal{E}_{1}\otimes\mathrm{Id}\right)\left(\rho\right)-\left(\mathcal{E}_{2}\otimes\mathrm{Id}\right)\left(\rho\right)\right|\right|_{1},
\end{equation}
where the trace norm
\begin{equation}
\left|\left|T\right|\right|_{1}\equiv\mathrm{Tr}\sqrt{T^{\dagger}T}.
\end{equation}
The diamond distance $D_{\diamond}\in\left[0,2\right]$ is related
to the minimal probability $p_{\mathrm{dist}}=\frac{1}{2}-\frac{1}{4}D_{\diamond}$ of distinguishing two channels by allowing quantum entangled states between the channel and an ancillary
space.

\begin{figure*}[hbt]
\centering \includegraphics[width=0.75\textwidth]{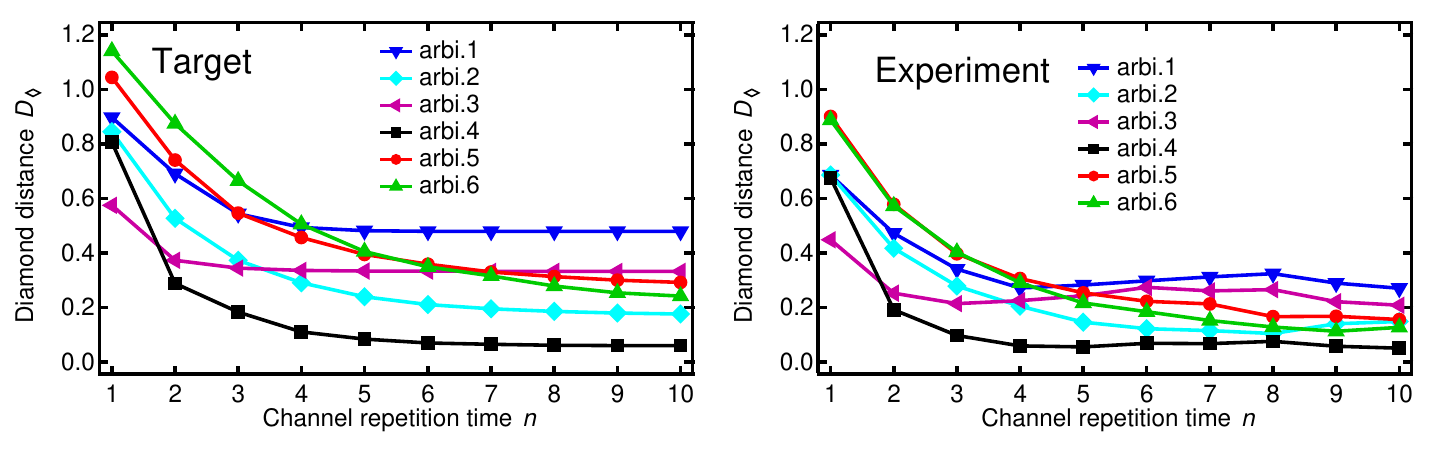} \caption{\textbf{Diamond distances $D_{\diamond}$ between the target channel $\chi_{\mathrm{T}}$ (experimental channel $\chi_{\mathrm{E}}$) and the depolarization channel with $p=\frac{3}{4}$}.}
\label{fig:SI-diamond}
\end{figure*}

In the main text, both $F_{\mathrm{G}}$ and $D_{\diamond}$ are shown
for different number of channel repetition $n$. It is shown that $F_{G}$
increases with $n$ while $D_{\diamond}$ decreases with $n$ for certain channels.
To explain this behavior, we also present $D_{\diamond}$
between the target channel $\chi_{\mathrm{T}}$ (experimental channel
$\chi_{\mathrm{E}}$) and the depolarization channel with $p=\frac{3}{4}$
\begin{equation}
\chi_{\mathrm{dpl}}=\frac{1}{4}\left[\begin{array}{cccc}
1 & 0 & 0 & 0\\
0 & 1 & 0 & 0\\
0 & 0 & 1 & 0\\
0 & 0 & 0 & 1
\end{array}\right].
\end{equation}
The results are presented in Fig.$\,$\ref{fig:SI-diamond}. These results indicate that repetitive arbitrary quantum channel eventually approaches
a depolarization channel, therefore the distance between the target channel and the experimental channel may be closer after several rounds of repetition.

\subsubsection{Correction of the measurement-induced phase shift}

\begin{figure*}[hbt]
\centering \includegraphics{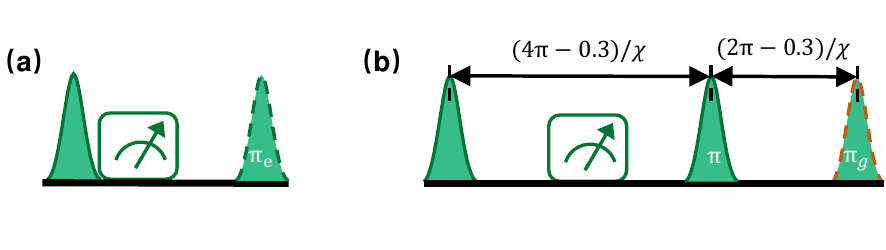} \caption{\textbf{Correction of the measurement-induced phase shift in the arbitrary
channel simulation.} The measurement-adaptive sequence (a) used in
both dephasing and damping channel simulations is replaced by the
new one in (b). The goal is to acquire integer multiples of 2$\pi$
phase for both $\ket{g}\ket{1}$ and $\ket{e}\ket{1}$. Here $0.3/\chi_{{\rm {s}}}$
is the time interval for $\ket{e}\ket{1}$ state under free evolution
to acquire a phase of $17^{\circ}$, which is the measurement-induced
phase on both $\ket{g}\ket{1}$ and $\ket{e}\ket{1}$.}
\label{fig:measure_arbi}
\end{figure*}

The measurement-induced phase on the photonic qubit does not commute
with other processes in the arbitrary channel simulation. So we cannot
ignore it during the experiment and deal with it in the final data
analysis as for the cases of the dephasing and damping channel simulations.
Instead, we experimentally eliminate this extra phase by modifying
the measurement-adaptive sequence, as shown in Fig.~\ref{fig:measure_arbi}.

In the arbitrary channel simulation [Fig.~4(a) of the main text], there
are two measurement-adaptive processes: one is to get the quantum
random number for deciding different branches and the other one is
used in each branch. In our experiment, we want to keep the time interval
between measurement and the following rotation pulse constant (320~ns)
to minimize extra ancilla decoherence during this time interval. However,
the measurement-adaptive sequence in Fig.~\ref{fig:measure_arbi}(a)
(used in both dephasing and damping channel simulations) will not
lead to the same phase on the photonic qubit when associated with $\ket{g}$
and $\ket{e}$ respectively. To solve this problem, the measurement-adaptive
sequence is replaced by Fig.~\ref{fig:measure_arbi}(b). The time interval
between the pre-rotation pulse on the ancilla and the measurement
is increased such that the time interval between the pre-rotation
and the $\pi$ pulse following the measurement is $(4\pi-0.3)/\chi_{{\rm {s}}}$.
Here $0.3/\chi_{{\rm {s}}}$ corresponds to $17^{\circ}$ rotation
of $\ket{1}$ state of the photonic qubit due to the cross Kerr between
the readout cavity and the photonic qubit, independent of the ancilla
state. If the measurement outcome of the ancilla state is $\ket{e}$,
the $\pi$ pulse brings the ancilla to $\ket{g}$ after acquiring
a total 4$\pi$ phase for $\ket{e}\ket{1}$ state (including the measurement-induced
phase). Here, in the joint state notation the letters represent the
ancilla states while the numbers correspond to the photonic qubit states.
On the other hand, if the measurement outcome of the ancilla state
is $\ket{g}$, the $\pi$ pulse after the measurement brings the ancilla
to $\ket{e}$ to acquire an extra phase during the following $(2\pi-0.3)/\chi_{{\rm {s}}}$
interval before a second conditional $\pi$ pulse flips the ancilla
back to $\ket{g}$. In the end, the original $\ket{g}\ket{1}$ also
acquires a 2$\pi$ phase.

\section{Discussions}

\subsection{The limitations of dephasing and damping rates}

The ancilla qubit facilitates all channel simulations. Consequently,
its intrinsic decoherence prevents an arbitrary fast dephasing or
damping rate, and eventually sets a limit on these rates. In this
part, we show how the ancilla decoherence affects the simulated dephasing
and damping rates in the corresponding channels.

\subsubsection{The dephasing channel}

First, we show that the dephasing of the ancilla will not affect the
dephasing channel. The dephasing channel relies on the dispersive
interaction between the ancilla and the photonic qubit. Explicitly,
the joint state $\ket{e}\ket{1}$ acquires a phase relative to other
three states $\ket{e}\ket{0}$, $\ket{g}\ket{1}$, and $\ket{g}\ket{0}$
with a rate equal to the dispersive interaction strength $\chi_{{\rm {s}}}$.
Before the projective measurement in the channel simulation, the dephasing
of the ancilla only changes the sign associated with $\ket{e}$ state,
which has no observable effect on the following projective measurement
and adaptive control. Therefore, the dephasing channel remains unaffected,
and we only need to consider thermal excitation and ancilla decay.

After a waiting time of $60~{\mathrm{\mu}}$s, the ancilla qubit can
be considered in a thermal equilibrium state with an $\ket{e}$ state
population $n_{{\rm {th}}}=0.87\%$ based on an independent calibration
experiment. For an initial $\ket{e}$ state, the following channel
simulation still works well except that the probability causing a
phase flip of the photonic qubit changes from ${\rm sin}^{2}(\theta/2)$
to ${\rm cos}^{2}(\theta/2)$ due to the exchange of $\ket{g}$ and
$\ket{e}$.

Between the $X_{\theta}$ gate and the adaptive $\pi$ pulse, random
decay of the ancilla qubit from $\ket{e}$ to $\ket{g}$ causes random
phase on $\ket{1}$ state instead of a complete phase flip. Therefore,
this will reduce the phase-flip probability of $\ket{1}$. On average,
we can treat this reduction factor as $1-p_{1}/2$ with $p_{1}=(3\pi/\chi_{{\rm {s}}})/T_{1}$,
the probability of having an ancilla decay. The ancilla upwards transition
probability from $\ket{g}$ to $\ket{e}$ is tiny during the short
channel simulation process and can be neglected. Combining the above
two effects, thermal excitation during $\tau_{0}$ and ancilla decay,
we get the final phase flip probability for the photonic qubit as a
function of the rotation angle in $X_{\theta}$:
\begin{equation}
p_{\phi}^{{\rm {lim}}}=(1-n_{{\rm {th}}}){\rm sin}^{2}(\theta/2)(1-p_{1}/2)+n_{{\rm {th}}}{\rm cos}^{2}(\theta/2)(1-p_{1}/2).
\end{equation}
Then we can get the upper limit of the external and controllable dephasing
rate in our channel simulation:
\begin{equation}
\Gamma_{\phi}^{{\rm lim}}=-\frac{{\rm ln}(1-2p_{\phi}^{{\rm {lim}}})}{\tau_{0}},
\end{equation}
which is used for Fig.~2(e) in the main text.

\subsubsection{The damping channel}

The ancilla dephasing will not affect the performance of the damping
channel, provided the initial state of the photonic qubit is at $\ket{1}$
state. This is because the ancilla dephasing during the the parity-type
protocol ($X_{\theta}$, $\pi/\chi_{{\rm {s}}}$, $X_{-\theta}$)
only flips the phase on $\ket{e}\ket{1}$, and this flip has no effect
on the following projective measurement and adaptive control. However,
if the initial state is a superposition of $\alpha\ket{0}+\beta\ket{1}$,
the ancilla dephasing will affect the resulting channel. Here, we
focus on the case of an initial $\ket{1}$ state of the photonic qubit,
since this is a direct reflection of the damping rate of this channel
[Fig.~3(b) of the main text]. Therefore, we only need to consider the
thermal excitation and energy decay of the ancilla.

Similar to the case of the damping channel, after a waiting time of
$60~{\mathrm{\mu}}$s, the ancilla qubit can be treated in a thermal
equilibrium state. For an initial $\ket{e}$ state of the ancilla,
the following channel simulation also works well except that the probability
causing the photonic qubit decay changes from ${\rm sin}^{2}\theta$
to ${\rm cos}^{2}\theta$ due to the exchange of $\ket{g}$ and $\ket{e}$.

During the parity-type protocol, random decay of the ancilla qubit
from $\ket{e}$ to $\ket{g}$ {[}with a probability of $p_{1}=(\pi/\chi_{{\rm {s}}})/T_{1}${]}
changes the photonic qubit damping probability to ${\rm sin}^{2}(\theta/2)$
since only the second $X_{-\theta}$ is effective on the ancilla.
The ancilla decay process could also happen with a probability $p_{2}$
during the measurement and the following waiting time. Then the adaptive
$\pi$ rotation will flip the ancilla to $\ket{e}$ and mess up with
the final GRAPE pulse to flip the photonic qubit from $\ket{1}$ to
$\ket{0}$. Since the photonic qubit state is unknown and we simply
treat this process causes a damping with a probability of 1/2 (completely
mixed photonic qubit state). These processes will reduce the probability
of damping the photonic qubit, and based on a probability calculation
we finally get the damping probability limit $\gamma^{{\rm {lim}}}$
the channel:

\begin{align}
\gamma^{{\rm {lim}}} & =(1-n_{{\rm {th}}})\{[1-p_{1}{\rm sin}^{2}(\theta/2)](1-p_{2}){\rm sin}^{2}\theta+[1-p_{1}{\rm sin}^{2}(\theta/2)]p_{2}/2+p_{1}{\rm sin}^{2}(\theta/2){\rm sin}^{2}(\theta/2)\}\\
 & +n_{{\rm {th}}}\{[1-p_{1}{\rm cos}^{2}(\theta/2)](1-p_{2}){\rm cos}^{2}\theta+[1-p_{1}{\rm cos}^{2}(\theta/2)]p_{2}/2+p_{1}{\rm cos}^{2}(\theta/2){\rm sin}^{2}(\theta/2)\}
\end{align}
Then we can get the upper limit of the external and controllable damping
rate in our channel simulation:
\begin{equation}
\Gamma_{1}^{{\rm lim}}=-\frac{{\rm ln}(1-\gamma^{{\rm {lim}}})}{\tau_{0}},
\end{equation}
which is used for Fig.~3(e) in the main text.

\subsection{Fit of process fidelity curves in the dephasing and damping channel
simulations}

\begin{figure*}[hbt]
\centering \includegraphics{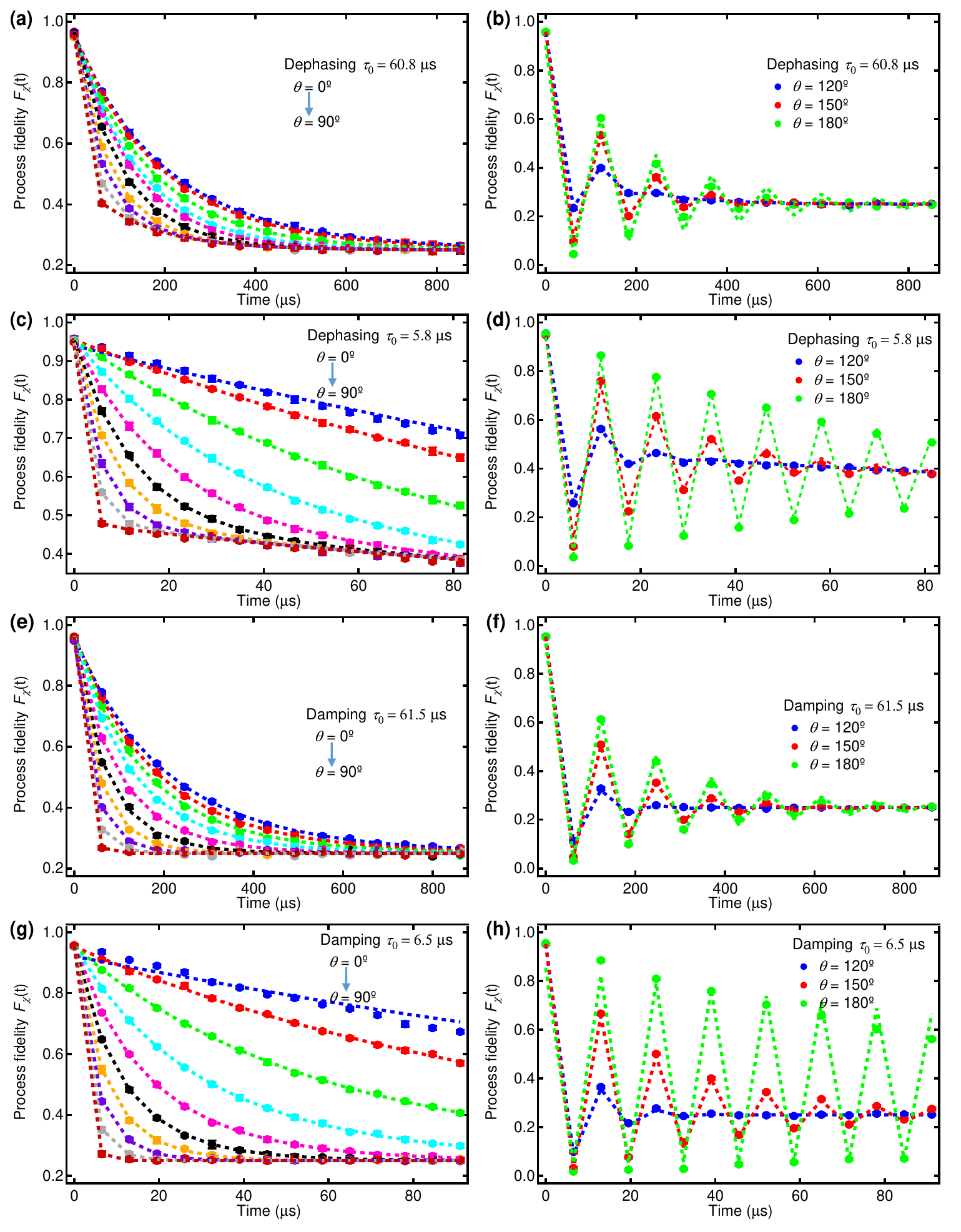} \caption{\textbf{Fit process fidelity curves with simulated $\chi$ matrices.}
We do numerical simulations with QuTip to compare with the experimental
data. To save time, we only numerically simulate the $\chi$ matrices
of two processes: free evolution of the system for $60~{\mathrm{\mu}}$s
and $5~{\mathrm{\mu}}$s respectively; and treat the channels as ideal
ones because of the large number of different channels. Due to the
repetitive nature of the channel, we then simply interleave the corresponding
free evolution process and the ideal channel (dephasing or damping)
for $n$ times to get the total process matrix $\chi(\theta)$ at different times. Finally, the experimental data are fitted with
$F_{\chi}=A[Tr(\chi_{{\rm I}}\chi(\theta)-0.25]+0.25$, where $\chi_{{\rm I}}$
is for an identity channel, $A$ is a pre-factor
to take into account the reduction due to the encoding and decoding
processes in the experiment but not in the simulation, and 0.25 is
final saturation value in the long time limit. The fitted results
are connected with dashed lines to guide the eye and are in excellent
agreement with the experiment.}
\label{fig:chi_fit}
\end{figure*}

The channel simulation results presented in the main text are obtained
with a waiting time $60~{\mathrm{\mu}}$s ($\tau_0 \approx 61~{\mathrm{\mu}}$s). To see the channel behavior
in a short time scale, we also perform simulations with a waiting time
$5~{\mathrm{\mu}}$s [Figs.~\ref{fig:chi_fit}(c) and (g)]. The channel
process fidelity, defined as the overlap of the measured $\chi$ matrix
and $\chi_{{\rm {I}}}$ for the identity operation, decays exponentially
when $\theta$ is small, but deviates from the exponential behavior
when $\theta$ is large. In addition, we perform the experiment with
$\theta>90^{\circ}$ as well, which corresponds to a phase-flip probability
$p_{\phi}>1/2$ for the dephasing channel [Fig.~\ref{fig:chi_fit}(d)]
or $-1\le\sqrt{1-\gamma}\le0$ for the damping channel [Fig.~\ref{fig:chi_fit}(h)].
In these two cases, the process fidelity oscillates after each extra
channel repetition.

\begin{figure}[hbt]
\centering \includegraphics{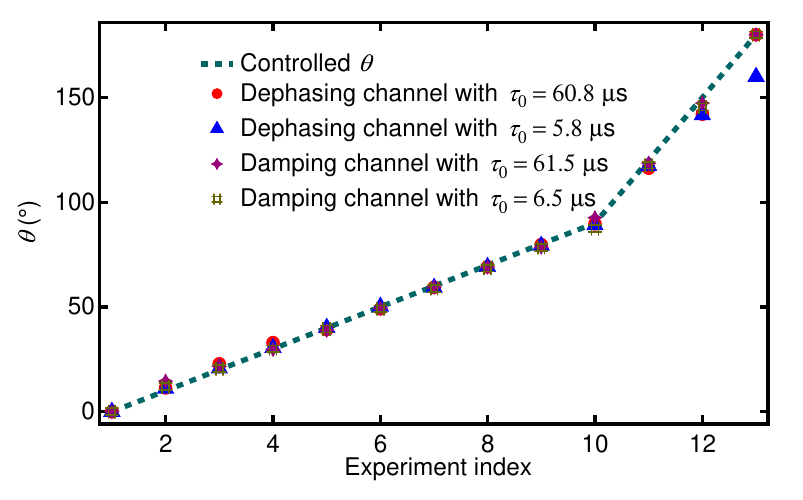} \caption{\textbf{$\theta$ extracted from fit of the process fidelity curves
in Fig.~\ref{fig:chi_fit}.} Points are the extracted $\theta$ (12
of them), in good agreement with the externally controlled rotation
angles in the experiment (connected with a dashed line).}
\label{fig:chi_fit_angle}
\end{figure}

In order to understand the decay and oscillation, and to verify the
channel performance, we can do full numerical simulations and compare
the results with the experimental data. To save time, we only numerically
simulate the $\chi$ matrices of two processes with QuTip in Python~\cite{Johansson2012,Johansson2013}:
free evolution of the system for $60~{\mathrm{\mu}}$s and $5~{\mathrm{\mu}}$s
respectively; and treat the channels as ideal ones because of the
large number of different channels. Then, for different input states
we can simply interleave the corresponding free evolution process
and the ideal channel (dephasing or damping) for $n$ times to get
the final states, and in turn the total process fidelity $\chi_{\theta}(t)$
at different times. We finally fit the experimental data with $F_{\chi}(t)=A[Tr(\chi_{{\rm I}}\chi_{\theta}(t)-0.25]+0.25$,
where $A$ is a pre-factor to take into account the reduction due
to the encoding and decoding processes in the experiment but not in
the simulation, and 0.25 is final saturation value in the long time
limit. The fitted results are shown in Fig.~\ref{fig:chi_fit} connected
with dashed lines to guide the eye, in excellent agreement with the
experiment. The extracted $\theta$'s are plotted in Fig.~\ref{fig:chi_fit_angle},
which also agree well with the externally controlled rotation angle
in the experiment.


\bibliographystyle{Zou}
%